\documentclass[MSNbibl,nameyear,dvips]{arxstspdf}
\usepackage{flushend}
\usepackage{stfloats}
\usepackage{graphicx,url,breakurl}

\volume{29}
\issue{3}
\pubyear{2014}
\firstpage{397}
\lastpage{419}
\doi{10.1214/13-STS465} 

\makeatletter
\renewcommand{\epsilon}{\varepsilon}
\renewcommand{\emptyset}{\varnothing}
\renewcommand{\citep}{\cite}
\renewcommand{\citealt}[1]{\citeauthor{#1}, \citeyear{#1}}
\newcommand{\xrightarrow}[1]{\stackrel{#1}{\longrightarrow}}
\makeatother

\begin{document}
\begin{frontmatter}

\title{Recursive Pathways to Marginal Likelihood Estimation with Prior-Sensitivity Analysis}
\runtitle{Recursive Pathways to Marginal Likelihood Estimation}

\begin{aug}
\author[a]{\fnms{Ewan} \snm{Cameron}\corref{}\ead[label=e1]{dr.ewan.cameron@gmail.com}}
\and
\author[a]{\fnms{Anthony} \snm{Pettitt}}

\runauthor{E. Cameron and A. Pettitt}

\affiliation{Queensland University of Technology}

\address[a]{Ewan Cameron is Research Associate, Science and Engineering Faculty, School of Mathematical Sciences
(Statistical Science),
Queensland University of Technology,
GPO Box 2434, Brisbane, QLD 4001, Australia \printead{e1}.
Anthony Pettitt is Professor,
Science and Engineering Faculty, School of Mathematical Sciences
(Statistical Science),
Queensland University of Technology,
GPO Box 2434, Brisbane, QLD 4001, Australia.}

\end{aug}

\begin{abstract}
We investigate the utility to computational Bayesian analyses of a particular
family of recursive marginal
likelihood estimators characterized by the (equivalent) algorithms known
as ``biased
sampling'' or ``reverse logistic regression'' in the statistics literature and ``the density of
states'' in physics.  Through a pair of numerical examples (including
mixture modeling of the well-known galaxy data set) we highlight
the remarkable diversity of sampling
schemes amenable to such recursive normalization, as well as the notable
efficiency of the resulting pseudo-mixture distributions
for gauging prior sensitivity in the Bayesian model selection context.  Our key
theoretical contributions are to introduce a novel heuristic (``thermodynamic integration via importance
sampling'') for qualifying the role of the bridging sequence in this
procedure and to reveal various connections
between these recursive estimators and the nested
sampling technique.
\end{abstract}


\end{frontmatter}

\section{Introduction}\label{intro}

Though typically unnecessary for computational parameter inference in
the Bayesian framework, the \mbox{factor}, $Z$, required to
normalize the
product of prior and likelihood nevertheless plays a vital role in
Bayesian model selection and model averaging
(Kass and Raftery, \citeyear{kas95}; \citep{hoe99}).  For priors admitting an  ``ordinary'' density,
$\pi(\theta)$, with respect to
the Lebesgue measure (a ``$\Lambda$-density''), we write for
the posterior
%
\begin{eqnarray}
\label{bayes} \pi(\theta|y)&=&L(y|\theta) \pi(\theta)/Z \quad\mbox{with}
\nonumber
\\[-8pt]
\\[-8pt]
 Z &=& \int
_{\Omega} L(y|\theta) \pi(\theta) \,d\theta,\nonumber
\end{eqnarray}
and, more generally (e.g., for stochastic process priors) we write
%
\begin{eqnarray}
\label{bayesalt}
dP_{\theta|y}(\theta)&=&L(y|\theta) \,dP_{\theta}(\theta)
/Z \quad\mbox{with}
\nonumber
\\[-8pt]
\\[-8pt]
 Z &=& \int_{\Omega} L(y|\theta) \bigl
\{dP_{\theta}(\theta)\bigr\},\nonumber
\end{eqnarray}
with the likelihood, $L(y|\theta)$, a non-negative,
real-valued function supposed integrable with respect to the prior.
In this context $Z$ is generally referred to as
either the \textit{marginal likelihood} (i.e., the likelihood of the observed
data marginalized [averaged] over the prior) or the
\textit{evidence}.  With the latter term though, one risks the
impression of
overstating the value of this statistic in the case of limited prior
knowledge (cf. \citealt{gel04}, Chapter~6).

Problematically, few complex statistical problems admit an analytical
solution to Equations (\ref{bayes}) or (\ref{bayesalt}), or span such low-dimensional spaces
[$D(\theta) \lesssim 5$--10] that direct numerical integration presents a viable alternative.  With errors
(at least in
principle) independent of dimension, Monte Carlo-based
integration methods have thus become the mode of choice for marginal likelihood estimation
across a diverse range of scientific disciplines, from evolutionary biology
(\citep{xie11}; \citep{ari12}; \citep{bae12}) and cosmology
(\citep{muk06}; \citep{kil10}) to quantitative finance (\citep{li11}) and sociology (\citep{cai12}).

\subsection{Monte Carlo-Based Integration Methods}

With the posterior most often ``thinner-tailed'' than the prior and/or
constrained within a much diminished sub-volume of the given parameter space, the simplest marginal
likelihood estimators drawing solely from
$\pi(\theta)$ or  $\pi(\theta|y)$ cannot be relied upon for
model selection purposes.  In the first case---strictly, that
$\int_\Omega [1/L(y|\theta)] \pi(\theta)\,d\theta$ diverges---the harmonic mean estimator
(HME; \citealt{new94}),
\[
\hat{Z}^H = \Biggl[ \sum_{i=1}^n
1/n/L(y|\theta_i) \Biggr]^{-1} \quad\mbox{for }
\theta_i \sim \pi(\theta|y),
\]
suffers \textit{theoretically} from an infinite variance, meaning
\textit{in practice} that its
convergence toward the true $Z$ as a
one-sided $\alpha$-stable limit law can be incredibly slow
(\citep{wol12}).  Even when ``robustified'' as per \citet{gel94}
or \citet{raf07}, however, the HME remains notably insensitive to
changes in $\pi(\theta)$, whereas $Z$ itself is characteristically
sensitive (\citep{rob09}; \citep{fri12}).  [See also \citet{wei12} for yet
another approach to robustifying the HME.]  Though assuredly finite
by default, the variance of the prior arithmetic mean estimator (AME),
\[
\hat{Z}^A = \sum_{i=1}^{n}
L(y|\theta_i)/n \quad\mbox{for } \theta_i \sim \pi(
\theta),
\]
on the other
hand, will remain impractically large whenever there exists a substantial difference in ``volume''
between the regions of greatest concentration in prior
and posterior mass, with huge sample sizes
necessary to achieve reasonable accuracy  (e.g., \citealt{nea99}).

A wealth of more sophisticated integration methods have thus lately
been developed for
generating improved estimates of the marginal likelihood, as reviewed
in depth by \citet{che00}, \citet{rob09} and \citet{fri12}. Notable
examples include the following: adaptive multiple importance sampling
(\citep{cor12}), \mbox{annealed} importance sampling (\citep{nea01}), bridge sampling
(\citep{men96}), [ordinary] importance sampling (cf. \citealt{liu01}), path
sampling/\allowbreak  thermodynamic integration (\citep{gel98}; \citep{lar06};
\citep{fri08};
\citep{cal09}), nested sampling
(\citep{ski06}; \citep{fer08}), nested importance sampling (\citep{cho10}),
reverse
logistic regression (\citep{gey94}), sequential Monte Carlo (SMC;
\citealt{cap04}; \citealt{del06}), the Savage--Dickey density ratio (\citep{mar10}) and the density of states (\citep{hab12}; \citep{tan12}).
A common thread running through almost all these schemes is the aim for a
superior exploration of the relevant parameter space via ``guided'' transitions
across a sequence of intermediate distributions, typically following a
bridging path between the
$\pi(\theta)$ and $\pi(\theta|y)$
extremes.  [Or, more generally, the $h(\theta)$ and $\pi(\theta|y)$
extremes if a suitable auxiliary/reference density, $h(\theta)$, is
available to facilitate the integration; cf. \citet{lef10}.]
However, the nature of this bridging path differs significantly
between algorithms.  Nested sampling, for instance,
evolves its ``live point set'' over a sequence of constrained-likelihood
distributions, $f(\theta)\propto\pi(\theta)I(L(y|\theta)\ge L_\mathrm{lim})$,
transitioning from the prior ($L_\mathrm{lim}=0$) through to the
vicinity of peak likelihood ($L_\mathrm{lim}\approx
L_\mathrm{max}-\epsilon$), while thermodynamic integration, on the
other hand, draws progressively (via Markov Chain Monte Carlo [MCMC];
\citealt{tie94}) from the family of ``power posteriors,''
%
\begin{equation}
\label{pp} \pi_t(\theta|y) \propto \pi(\theta)L(y|
\theta)^t,
\end{equation}
explicitly connecting the prior at $t=0$ to the posterior at
$t=1$.

Another key point of comparison between these rival Monte Carlo
techniques lies in their choice of identity by which the evidence is
ultimately computed.  The (geometric) path sampling
identity,
\[
\log Z = \int_0^1 \mathrm{E}_{\pi_t}
\bigl\{ \log L(y|\theta) \bigr\} \,dt,
\]
for example, is shared across both thermodynamic integration and SMC,
in addition to its namesake.  However, SMC can also be run with the ``stepping-stone'' solution (cf. \citealt{xie11}),
\[
Z = \prod_{j=2}^m Z_{t_j}/Z_{t_{j-1}},
\quad\mbox{where } t_1=0 \mbox{ and } t_m=1,
\]
with $\{t_j\dvtx  j=1,\ldots, m\}$ indexing a sequence of (``tempered'') bridging
densities, and, indeed, this is the mode preferred by experienced practitioners
(e.g., \citealt{del06}).  Yet another identity for computing the marginal likelihood is that
of the recursive pathway explored here.

First introduced within the ``biased sampling'' para\-digm
(\citep{var85}), the recursive pathway is shared by the popular
techniques of ``reverse logistic regression'' (RLR) and ``the density of
states'' (DoS).  By \textit{recursive} we mean that, algorithmically,
each may be run such that the desired $Z$ is obtained through
backward induction of the complete set of intermediate
normalizing constants corresponding to the sequence of distributions in the given bridging path
by supposing these
to be already
known.  That is, a stable solution may be found in a
Gauss--Seidel-type manner (\citep{ort67}) by
starting with a guess for each normalizing constant as input to a convex system
of equations for updating these guesses, returning the new output
as input to the same equations, and iterating until convergence.  In fact, although the RLR and the DoS approaches differ
vastly in concept and derivation---the former emerging from
considerations of the reweighting mixtures problem in applied statistics
(\citep{gey92}; \citep{gey94}; \citep{che97}; \citep{kon03}) and the latter from computational strategies for free
energy estimation in physics/chemistry/biology
(\citep{fer89}; \citep{kum92}; \citep{shi08}; \citep{hab12}; \citep{tan12})---both may be seen to
recover the same algorithmic form in practice.  To illustrate
this equivalence, and to explain further the recursive pathway to
marginal likelihood estimation, we describe each in detail below
(Sections~\ref{rlrsection} and~\ref{dossection}),
though we begin with the more general biased sampling algorithm
(Section~\ref{bs}).

Following this review of the recursive family (which includes our
theoretical contributions concerning the link between the DoS and
nested sampling in Section~\ref{rns}), we highlight the potential for
efficient prior-sensitivity analysis when using these marginal
likelihood estimators (Section~\ref{psa}) and discuss issues regarding
design and sampling of the bridging sequence (Section~\ref{fisher}).
We then introduce a novel heuristic to help inform the latter by
characterizing the connection between the bridging sequences of biased
sampling and thermodynamic integration (Section~\ref{tivis}).
Finally, we present two numerical case studies illustrating the issues
and techniques discussed in the previous sections: the first concerns
a ``mock'' banana-shaped likelihood function (Section~\ref{banana}) and
includes the demonstration of a novel synthesis of the recursive
pathway with nested sampling (Section~\ref{nestedcomp}), while the
second concerns mixture modeling of the galaxy data set (Section~\ref{mixtures}) and\vadjust{\goodbreak} includes a demonstration of importance sample
reweighting of an infinite-dimensional mixture posterior to recover
its finite-dimensional counterparts (Section~\ref{immr}).

\section{Biased Sampling}\label{bs}

The archetypal recursive marginal
likelihood esti\-mator---from which both the RLR and DoS methods may be
directly recovered---is that of biased sampling,
introduced by \citet{var85} for finite-dimensional parameter spaces
and extended to general sample spaces by \citet{gil88}.  The
basic premise of biased sampling is that one has available $m$ sets of $n_j$
i.i.d. draws, $\{\theta_i\}_j$, from a series of $w_j(\theta)$-weighted
versions of a common, unknown measure, $F$, that is,
\[
\{\theta_i\}_j \sim F_j, \quad\mbox{where } dF_j(\theta) = w_j(\theta)/W_j \,dF(
\theta).
\]
The $W_j$ term here represents the
normalization constant of the $j$th weighted distribution, typically unknown.  As \citet{var85} demonstrates,
provided the drawn $\{\theta_i\}_j$ obey a certain graphical condition
(discussed later), then there exists a unique nonparametric maximum
likelihood estimator (NPMLE) for $F$, which as a by-product produces
consistent estimates of all unknown $W_j$.  If the common measure, $F$, is in fact the parameter prior,
$P_\theta$, then the choices $w_1(\theta) = 1$ and $w_m(\theta) =
L(y|\theta)$ describe sampling from the prior and posterior,
respectively.  Hence, we switch to the notation $W_j = Z_j$ with $Z_1
= 1$ (for a proper prior) and $Z_m = Z$ for the above choices of $w_1$ and $w_m$.

For a given bridging scheme to be amenable to normalization via biased
sampling, it is of course necessary that each intermediate sampling
distribution be absolutely continuous with respect to the prior (i.e.,
$P_j \ll P_\theta$) such that the weight function corresponds to the
Radon--Nikodym derivative,
$w_j(\theta)=\frac{dP_j}{dP_\theta}(\theta)$.  It is easy to verify
then the applicability of biased sampling to, for example, \textsc{(i)} importance
sampling from a sequence of bridging densities, $f_j(\theta)$, with
(at least the union of their) supports matching but not exceeding
that of a $\Lambda$-density prior, $w_j(\theta) =
f_j(\theta)/\pi(\theta)$; and \textsc{(ii)}~thermodynamic integration
over tempered likelihoods, $w_j(\theta)=L(y|\theta)^{\beta_j}$,
for both the $\Lambda$-density and general case.  In fact, if we view the likelihood function as
defining a transformation of the prior, $P_\theta$, to the measure $P_L$ in univariate ``likelihood
space,'' $0 \le L \le \infty$, then such tempering
may be seen as directly analogous to Vardi's example of ``length biased sampling.''  Accordingly, Vardi's case study of $m=2$ with $w_1 = 1$ and $w_2 =
x$ (read $L$) equates to marginal
likelihood estimation via defensive
importance sampling from the prior and posterior (\citep{new94}; \citep{hes91}),
while his one sample study with $w_1 = x$ ($L$)
matches the HME.

For Bayesian analysis problems in which the prior measure is explicitly known (as
opposed to being ``known'' only implicitly as the induced measure belonging to a
well-defined stochastic process), the application of the biased sampling paradigm to the task of
marginal likelihood estimation is arguably paradoxical since we make
the pretence to estimate $P_\theta$ (known) in order to recover
an estimate for $Z$ (unknown).  However, we would propose that
an adequate justification for the use of Vardi's method in this
context is already
provided by
the same pragmatic reasoning used to adopt \textit{any}
statistical estimator for the task of marginal likelihood computation
in place of the direct approach of numerical integration (quadrature)---namely,
that although $Z$ is defined exactly by our known prior and
likelihood function, we choose to treat it as if it were an unknown variable simply
because the MC integration techniques this brings into play
are more computationally efficient (being relatively
insensitive to the dimension of the problem; cf. \cite{liu01}).

Vardi's derivation of the NMPLE for the unknown $F$ (i.e., $P_\theta$) in
biased sampling involves two key steps.  The first is the observation
that, as is typical of the NMPLE method in general, the resulting estimator,
$d\hat{F}(\theta)$, will be strictly atomic with point masses assigned to each of the
sampled $\theta_i$ (also called a histogram estimate of $F$).  The
second is that the normalization constants for each $W_j$
corresponding to the atomic $d\hat{F}(\theta)$ can then be
learned via an appropriately
weighted summation over \textit{all} the observed $\theta_i$ (not just
those from the corresponding $j$th distribution).  In the notation
for our marginal likelihood estimation scenario, \citet{var85} shows
that the estimation problem for $d\hat{P}_\theta(\theta) = \{p_i\}_j$ can ultimately be reduced to the
maximization of the following log-likelihood function,
\[
\log \mathcal{L}(p) = \sum_{j=1}^m \sum
_{i=1}^{n_j} \log \bigl( w_j
\bigl(\{\theta_i\}_j\bigr) \{p_i
\}_j / \hat{Z}_j \bigr),
\]
subject to the constraints, $\sum_{j=1}^m \sum_{i=1}^{n_j} \{p_i\}_j =
1$ and all $\{p_i\}_j > 0$ [see Vardi's Equation (2.2), where we avoid his explicit
treatment of matching $\theta_i$ draws, implicitly allowing multiple point mass contributions at the same $\theta_i$ to
give a summed contribution to the atomic $d\hat{P}_\theta (\theta)$].\vadjust{\goodbreak}

Importantly, the resulting biased sampling estimator for the unknown $Z_k$
allows for a recursive solution via the iterative updating of initial
guesses ($\hat{Z}_k > 0$) as follows:
%
\begin{eqnarray}
\label{Zupdates} \hat{Z}_k &=& \sum_{j=1}^m
\sum_{i=1}^{n_j} \Biggl( w_k
\bigl(\{\theta_i\}_j\bigr)
\nonumber
\\[-8pt]
\\[-8pt]
&&\phantom{\sum_{j=1}^m
\sum_{i=1}^{n_j} \Biggl(}{}\Big/ \Biggl(\sum
_{s=1}^m n_s w_s\bigl(\{
\theta_i\}_j\bigr)/ \hat{Z}_s\Biggr)
\Biggr)\nonumber
\end{eqnarray}
(adapted from
\citeauthor{gil88}'s \citeyear{gil88} Proposition~1.1c).  As discussed by
\citet{var85} and \citet{gey94}, the above system of $(m-1)$ equations in
$(m-1)$ unknowns (given $Z_1=1$) with Gauss--Seidel type iterative updates is globally convergent, although the gradient and Hessian of the
likelihood function are also accessible, meaning that alternative
maximization strategies harnessing this information may prove more
efficient within a restricted domain.

The convergence properties of the biased sampling estimator for the unknown $F$
(i.e., $P_\theta$) and its associated $W_j$ ($Z_j$) in general state spaces
(possibly infinite-dimensional) have been thoroughly characterized by
\citet{gil88} using the theory of empirical processes indexed by
sets and functions (cf. \citealt{dud83}).  In particular,
\citet{gil88} demonstrate a central limit theorem (CLT) for convergence of the
vector of normalization estimates, $\hat{\mathbf{W}}$, to the
truth, $\mathbf{W}$, as $\sqrt{N}(\hat{\mathbf{W}} - \mathbf{W}) \xrightarrow{d}
\mathcal{N}(\mathbf{0},\Sigma)$, where the covariance matrix, $\Sigma$,
takes the form given in their Proposition~2.3 [for the case here of
$Z_1=1$ known, otherwise their Equation (2.24)].  The sample-based
estimate of this error matrix, $\hat{\Sigma}$, is easily computed from
the output of a standard biased sampling simulation, and in our numerical
experiments with the banana-shaped pseudo-likelihood function of
Section~\ref{banana} it was observed to give (on average, with an approximate transformation via Slutsky's
lemma) a satisfactory, though slightly conservative,
match to the sample variance of $\log \hat{Z}$ under repeat
simulation, even at relatively small sample sizes.

However, as noted by Christian Robert in his discussion of \citeauthor{kon03}'s (\citeyear{kon03})
``read'' paper, the
availability of such formulae (for the asymptotic
covariance matrix) can sometimes
``give a false confidence in \mbox{estimators} that should not be used.''
A~canonical example is that of the HME, for which the usual importance sampling variance formula\vadjust{\goodbreak}
applied to the posterior draws may well give a finite result, though in fact the
theoretical variance is infinite (meaning
that the convergence of the HME is no longer obeying the assumed
CLT).  In particular, for finite theoretical
variance of the HME (cf. Section~\ref{intro}) we require that the prior
is fatter tailed than the posterior such that $\int_\Omega
[1/L(y|\theta)]\pi(\theta)\,d\theta < \infty$.  As was recognized by
\citet{var85} and \citet{gil88}, the same condition effectively
holds for the validity of the CLT for biased sampling and may
be expressed as an inverse mean bias-weighted integrability requirement
over the indexing class of functions or sets in
its empirical process construction.  Important to note in the context of
marginal likelihood estimation is that
provided the prior itself is contained within the weighting scheme
[e.g., $w_1(\theta)=1$], then the above condition
is automatically satisfied; this of course parallels the strategy of defensive
importance sampling (\citep{new94}; \citep{hes91}).

Finally, we observe here the other key prerequisite for
successful biased sampling: that the bridging sequence of weighting functions and
the random draws from them are such that a unique NPMLE for $F$ ($P_\theta$)
actually exists.  To ensure the \textit{asymptotic} existence of a unique NPMLE (i.e.,
with an unlimited number of draws from each weighted distribution),
\citet{var85} gives the following condition on the supports, $\operatorname{Supp}(w_j)$, of the
bridging sequence: that there does not exist a proper
subset, $B$, of $\{1,\ldots,m\}$ such that
\[
\biggl(\bigcup_{j\in B}\operatorname{Supp}(w_j)
\biggr) \cap \biggl(\bigcup_{j\notin
B}\operatorname{Supp}(w_j)
\biggr) = \emptyset.
\]
In effect, the set of bridging distributions must overlap in
such a way that the relative normalization of each with respect to
all others will be inevitably
constrained by the data.  This condition is again satisfied
automatically if the support of at least one of the bridging distributions encompasses all
others, such as that of the prior or an equivalent reference
density. In the finite sample sizes of real-world simulation the above
must be strengthened to specify that the drawn $\{\theta_i\}_j$ do in fact
cover each  critical region of overlap.  Formally,
\citet{var85} introduces a requirement of strong connectivity on the directed graph,
$G$, with $m$ vertices and edges $h$ to $j$ for each $(h,j)$-pairing, such that
$w_h(\theta_k) > 0$ for some $\theta_k \in \{\theta_i\}_j$.
This is equivalent to the finite sample ``inseparability'' condition
given by \citet{gey94}.

\subsection{Reverse Logistic Regression}\label{rlrsection}

In the reweighting mixtures problem (cf.
Geyer and Thompson, \citeyear{gey92} and \citealt{gey94}) the aim is to discover an
efficient
proposal density for use in the importance sampling
of an arbitrary target about which little is known {a priori}.
Geyer's solution was to suggest sampling not from a single density of
standard form,
but rather from
an \textit{ensemble} of different densities, $f_{j}(\theta) =
q_{j}(\theta)/Q_{j}$, for $j=1,\ldots,m$ with
$q_{j}(\theta)$ known and $Q_{j}$ typically unknown.  The pooled draws,
$\{\{\theta_i\}_j\dvtx i=1,\ldots,n_j; j=1,\ldots,m\}$, are then to be treated as if
from a single mixture density, with each free normalizing constant---and hence
the appropriate weighting scheme---to be derived recursively.  As with
biased sampling, if we suppose $q_1(\theta)$ to be the Bayesian prior (with
$Q_1=1$) and $q_m(\theta)$ the (unnormalized) posterior (with $Q_j=Z$), the relevance
of this approach to marginal likelihood estimation becomes readily
apparent.  In this context we write the imagined (i.e., pseudo-) mixture density,
$p(\theta)$, in the form
%
\begin{equation}
\label{mixtured} p(\theta) = \sum_{j=1}^m
[n_j/n] \bigl[ q_{j}(\theta)/Z_j\bigr],
\end{equation}
where $n=\sum_{j=1}^m n_j$.\vspace*{1pt}

The recursive normalization scheme introduced by \citet{gey94} for this purpose is based on
maximization in $\{Z_2,\ldots,Z_m\}$ (i.e., $[\mathbb{R}_+]^{m-1}$) of the following \textit{quasi}-log-likelihood function
representing the likelihood of each set of $\{\theta_i\}_j$ having
been drawn from its true $f_{j}(\theta)$ rather than some other
$f_{k[\neq j]}(\theta)$ in the pseudo-mixture:
%
\begin{eqnarray}
\label{logLx} &&\log L\bigl(\bigl\{\{\theta_i\}_j\dvtx
i=1,\ldots,n_j; j=1,\ldots,m\bigr\}|\nonumber
\\
&&\hspace*{128pt}\{Z_1,
\ldots,Z_m\}\bigr)
\\
&&\quad= \sum_{j=1}^m \sum
_{i=1}^{n_j} \log \bigl( q_{j}\bigl(\{
\theta_i\}_j\bigr)/Z_j /p\bigl(\{
\theta_i\}_j\bigr) \bigr).
\nonumber
\end{eqnarray}
Owing to the arithmetic equivalence between Equation (\ref{logLx}) and
the objective function of logistic regression in the generalized
linear modeling framework---but with the ``predictor'' here random
and the ``response'' nonrandom---\citet{gey94} has dubbed this method
``reverse logistic regression.'' Setting the partial derivative in each unknown $Z_{k}$ to zero
yields the series of convex equations defining the RLR marginal likelihood estimator:
%
\begin{equation}
\label{eleven} \hat{Z}_{k} = \sum_{j=1}^m
\sum_{i=1}^{n_j} q_{k}\bigl(\{
\theta_i\}_j\bigr)/p\bigl(\{\theta_i
\}_j\bigr)/n,
\end{equation}
which, with reference to our definition of the
pseudo-mixture density above, may be confirmed equivalent to
biased sampling [Equation (\ref{Zupdates})]
in the $\Lambda$-density case for $w_j(\theta) =
q_j(\theta)/\pi(\theta)$. [The $\pi(\theta)$ term ultimately
cancels out from both the numerator and denominator of Equation
(\ref{Zupdates}), but serves here to establish our connection with the
notion of a common unknown distribution, $F$ or $P_\theta$.]

As \citet{kon03} explore in detail, the fact that Geyer's RLR
derivation via the quasi-log-likelihood function of Equation (\ref{logLx})
leads to the same set of recursive update equations as Vardi's
biased sampling hides a certain weakness of this ``retrospective
formulation'': that the Hessian of the \textit{quasi}-log-likelihood
does not provide the correct asymptotic covariance matrix for the
output $\hat{Z}_k$.  (Though the difference in practice is almost
negligible; cf. Section~\ref{banana}.) The same applies to a
``na\"ive,'' alternative derivation of the RLR estimator---relevant to the
thermodynamic integration via importance sampling methodology we
describe in Section~\ref{tivis}---given by \citet{jia03} in
their discussion of Kong et al.'s ``read'' paper.  That is,
treat the pooled $\{\theta_i\}_j$ as if drawn from the
pseudo-mixture density, $p(\theta)$, with $Z_k$ ($k=2,\ldots,m$)
unknown, and apply the ordinary importance sampling
estimator---based on the identity, $Z_{k} = \int_{\Omega}
\frac{q_{k}(\theta)}{p(\theta)} p(\theta)\,d\theta$---to recover the
recursive update scheme of Equation
(\ref{Zupdates}) (but again without a corresponding argument to arrive
at the correct variance).

An interesting observation often made in connection with RLR is that
Equation (\ref{eleven}) can in fact be applied
without knowledge of which $f_j(\theta)$ each $\theta_i$ was drawn from,
such that we may rewrite the recursive update scheme,
%
\begin{equation}
\label{rlrx} \hat{Z}_{k} = \sum_{i=1}^{n}
q_{k}(\theta_i)/p(\theta_i)/n,
\end{equation}
where we have taken the step of ``losing the labels,'' $j$, on our
$\{\theta_i\}_j$.  This is made possible, as \citet{kon03} explain,
because ``under the model as specified \ldots  the
association of draws with distribution labels is uninformative.  The
reason for this is that all the information in the labels for
estimating the ratios is contained in the design constants,
$\{n_1,\ldots,n_m\}$.''

\subsection{The Density of States}\label{dossection}

Yet another construction of the convex series of $\hat{Z}_{k}$ updates
characterizing the recursive appoach [cf. Equation (\ref{Zupdates})] has recently been
demonstrated in the context of free
energy estimation for molecular interactions by \citet{hab12} and
\citet{tan12}.  In this framework rather than aiming directly for
estimation of the marginal likelihood one aims instead to
reconstruct a closely-related distribution, namely, ``the density of states''
(DoS), $g(e)$, defined in the physics literature in terms of a
composition of the Dirac delta ``function,'' $\delta(\cdot)$, as
\[
g(e) = \int_\Omega \pi(\theta)\delta\bigl(e+\log L(y|\theta)
\bigr) \,d\theta.
\]
Important to note from a mathematical perspective, however, is that the composition of the Dirac delta
``function''---which is itself not strictly a function, being definable
only as a measure or a generalized function---lacks an
intrinsic definition.  \citet{hor83} proposes a version in $\mathbb{R}^n$
valid
only when the composing function, here
$v(\theta)=e+\log
L(y|\theta)$, is continuously differentiable and $dv(\theta)/d\theta$
nowhere zero, clearly problematic whenever
the likelihood function holds constant over a set of nonzero measures with respect
to $P_\theta$!  We therefore begin by suggesting a robust, alternative
definition of the DoS as a transformation of the likelihood through
the prior, an exercise that also serves to elucidate its connections
with Skilling's nested sampling.

As briefly noted earlier with respect to characterization of the HME
as
Vardi's ``length biased sampling,'' the
likelihood function can serve as the basis for construction of a
number of measure theoretic
transformations of the prior.   Most notably, the mapping
$L(y|\theta)\dvtx  \Omega \mapsto \mathbb{R}^+$ gives the prior in likelihood space ($0 \le L \le
\infty)$,
\[
P_L\dvtx \mathrm{P}_L\{B\}= \int_{L^{-1}B}
\bigl\{dP_\theta(\theta)\bigr\}
\]
for $B \in \mathcal{B}(\mathbb{R}^+)$ (the Borel sets on the extended
reals) following \citeauthor{hal50} [(\citeyear{hal50}), page 163], with the notation $L^{-1}B$ denoting
the (assumed $P_\theta$-measurable) set of all $\theta$ transformed through $L(y|\theta)$ into $B$.  If
the domain of $\theta$ is a metric space, then continuity (or
at least discontinuity on no more than a countable set) of
$L(y|\theta)$ is sufficient to ensure the $P_\theta$-measurability of $B$ (i.e., the validity of the
above), while the continuity of the logarithm in $e(\theta)=-\log
L(y|\theta)$ ensures the same for the corresponding transformation of
the prior to ``energy'' space ($-\infty \le e \le \infty$),
\[
P_e\dvtx \mathrm{P}_e\{C\}= \int_{e^{-1}C}
\bigl\{dP_\theta(\theta)\bigr\},
\]
with $C \in \mathcal{B}(\mathbb{R}^+)$. In each case the appropriate
version of the marginal likelihood shares equality with the
original [Equation (\ref{bayesalt})] wherever $Z$ is itself finite,
owing to the $P_L$- and $P_e$-measurability of $L$ and $\exp(-e)$,
respectively:
\begin{eqnarray*}
Z &=& \int_0^\infty L \bigl\{dP_L(L)
\bigr\} \quad\mbox{and}
\\
 Z &=& \int_{-\infty}^\infty
\exp(-e) \bigl\{dP_e(e)\bigr\}
\end{eqnarray*}
(cf. \citealt{hal50}, page 164).

Although unnecessary for a straightforward application of biased sampling, one might
choose to further require that $P_e$ admit a $\Lambda$-density,
equivalent to the requirement that its distribution function,
$G_e(e^\prime) = \int_{-\infty}^{e^\prime} \{dP_e(e)\}$, be everywhere
differentiable.  For a continuous likelihood function we can be assured
of this provided that $L(y|\theta)$ at no place
holds constant over a set of nonzero measures with respect
to $P_\theta$---the same limitation on its $\delta$ ``function''
definition.  If so, we may write the marginal likelihood integral
as \citet{hab12},
%
\begin{equation}
\label{dose} Z = \int_{-\infty}^\infty \exp(-e)
g_e(e) \,de.
\end{equation}

Estimation of $g_e(e)$ (or in fact the general measure, $P_e$) can of
course be accomplished via biased sampling given i.i.d.'s draws
from a series of $w(e)$-weighted versions of $g_e$, and, indeed, this
is the justification of the DoS algorithm---seen as the limiting case of
the weighted histogram analysis method (\citep{fer89}) with bin size
approaching zero---given by \citet{tan12}.  The derivation of the
recursive update formula [Equation (\ref{Zupdates})] presented by
\citet{hab12} for the DoS is alternatively via a novel functional analysis
procedure for optimization of the log-likelihood of an empirical
energy histogram; however, as with Geyer's RLR derivation, this approach does not lead to an
uncertainty estimate or CLT for the output $\hat{Z}_k$.

\subsubsection{Relation to nested sampling}\label{rns}

The nested sampling
identity (\citep{ski06}),
%
\begin{equation}
\label{ns} Z = \int_0^1 L(X) \,dX,
\end{equation}
where $L(X)$ represents the inverse of the survival function of
likelihood with respect to the prior---that is, $X(L^\prime) = 1-\mathrm{P}_L\{L
\le L^\prime\}$---and $dX$ denotes Riemann integration over the ``prior
mass cumulant,'' may best be
understood by reference to a
well-known relation between the expectation of a non-negative random
variable and its distribution function, namely, that for $y \sim P_Y$ with
$y \ge 0$,
\[
\mathrm{E}_Y\{Y\} = \int_0^\infty Y
\bigl\{dP_Y(Y)\bigr\} = \int_0^\infty
\bigl( 1-\mathrm{P}_Y\{Y \le y\}\bigr) \,dy
\]
(cf.  \citealt{bil68}, page 223).  Importantly, this relation (which follows
from integration by parts) holds irrespective of whether or not $P_Y$
admits a $\Lambda$-density, and in the marginal likelihood context becomes $Z
= \int_0^\infty 1-\mathrm{P}_L\{L \le
L^\prime\} \,dL^\prime$. If $\mathrm{P}_L\{L=\infty\}=0$, then this
monotonically decreasing, \textit{cadlag} function on $\mathbb{R}_+$ with bounded range (between zero
and one) is (perhaps improper) Riemann integrable, and we may simply ``switch axes'' to
obtain Equation (\ref{ns}).
While the uniqueness of the inverse survival function, $L(X)$, can be
ensured by requiring $L(y|\theta)$ to be continuous with connected
support (\citep{cho10}), the weaker condition of $L(y|\theta)$
discontinuous on a set of measure zero with respect to
$P_L$ suffices to ensure an $L(X)$ defined uniquely on all but a
corresponding set of Lebesgue measure zero, negligible also for our
Riemann integration.

Now for differentiable $G_e(e^\prime) = \mathrm{P}_\theta\{e(\theta) < e^\prime\}$,
such that $g(e)$ might be defined without our earlier measure
theoretic considerations as $g(e) = dG_e(e)/de$, the DoS version of
the marginal likelihood
[Equation (\ref{dose})] can nevertheless be recovered using the nested
sampling identity.  Observing that $G_e(e) = X(\exp[-e]) = X(L)$,
we have
\begin{eqnarray*}
g(e)&=&dX\bigl(\exp[-e]\bigr)/de = dX(L)/dL \times dL/de
\\
&=& dX(L)/dL \times -
\exp[-e].
\end{eqnarray*}
Substitution of $X(L)$ into
Equation (\ref{ns}) yields
\begin{eqnarray*}
Z &=& \int_{X(\infty)}^{X(0)}L(X)\,dX
\\
&=& \int
_\infty^0 L\bigl(X\bigl(L^\prime\bigr)\bigr)
\times dX\bigl(L^\prime\bigr)/dL^\prime \times dL^\prime,
\end{eqnarray*}
and then by substitution of $e$ we recover
\begin{eqnarray*}
Z&=& \int_{e(-\infty)}^{e(\infty)} L^\prime \times dX
\bigl(L^\prime\bigr)/dL^\prime \times dL^\prime
\\
&=&\int
_{-\infty}^{\infty}\exp[-e]\times -g(e)\exp[e]\times -
\exp[-e] \times de.
\end{eqnarray*}
That is, consistent with the requirements of \citet{hab12} and
\citet{tan12}, this alternative DoS formulation returns the identity
\[
Z=\int_{-\infty}^{\infty}g(e)\exp(-e)\,de.
\]

Interestingly, the above relationship between the DoS and nested
sampling identities is mirrored by the existence of a measure
theoretic construction for the latter (cf. Appendix C of
\citealt{fer13}).  If we take the survival function,
$X(L) = 1-\int_0^{L}\{dP_L\}$, as defining yet another
transformation of the
prior through the likelihood---a transformation ensured $P_L$-measurable,
and hence $P_\theta$-measurable, by the right continuity of $X(L)$---we recover the following
distribution in prior cumulant space ($0 \le X \le 1$):
\[
P_X\dvtx \mathrm{P}_X\{D\}= \int_{X^{-1}D}
\bigl\{dP_\theta(\theta)\bigr\}.
\]
Similarly, the marginal likelihood formula equivalent to the nested
sampling identity becomes
\[
Z = \int_0^1 L(X)\{dX\}
\]
for $X(L)$ invertible, that is, $L(y|\theta)$ continuous with connected
support (\citep{cho10}).  More generally, though, we can view
$L(X)$ as the conditional probability function of likelihood
given prior mass cumulant defined modulo $P_\theta$ by the relation
%
\begin{equation}
\label{nsx}  \int_{X^{-1}D}L(y|\theta)\bigl\{dP_\theta(
\theta)\bigr\} = \int_D e_\theta(L|X) \bigl
\{dP_X(X)\bigr\}\hspace*{-28pt}
\end{equation}
(cf.
\citealt{hal49}).  For statistical problems on a complete
separable metric space there will always exist a unique local version of
$e_\theta(L|X)$ defined as a weak limit such that $e_\theta(L|X=x)$
is meaningful even for atomic $x$ (\citep{pfa79}).

The value of this insight becomes apparent when we examine the nested sampling estimator
for posterior functionals (cf. \citealt{cho10}),
\[
\mathrm{E}_{\pi(\theta|y)}\bigl\{f(\theta)\bigr\} \approx \sum
_{i=1}^n \tilde{w}_i L(
\theta_i|y)f(\theta_i),
\]
where $\tilde{w}_i$ here represents the nested sampling posterior
weight for $\theta_i$, $d\hat{P}_X(X(\theta_i))$---typically $\tilde{w}_i =
(\hat{X}_{i-1}-\hat{X}_i)$ (\citep{ski06}).  This estimator relies on
the relation given by Equation (\ref{nsx}) with $L(y|\theta)$ replaced
by $L(y|\theta)f(\theta)$, which holds for $f(\theta)$
measurable---a~more general condition than that of $e_\theta(f|L)$
absolutely continuous
given by \citet{cho10}.  Importantly, this ensures the validity of
prior-sensitivity analysis via computation of the posterior
functional of $\pi_\mathrm{alt}(\theta)/\pi(\theta)$ in nested
sampling---a powerful technique not previously exploited in nested
sampling analyses---as we shall discuss for the case of biased sampling below.

\subsection{Importance Sample Reweighting for Prior-Sensitivity Analysis}\label{psa}

In the Bayesian framework (\citep{jef61}; \citep{jay03}) the ratio of marginal likelihoods
under rival hypotheses (i.e., the Bayes factor) operates directly on
the prior odds ratio for model selection to produce the posterior
odds ratio as
%
\begin{eqnarray}
&&\mathrm{P}\{M_1|y\}/\mathrm{P}\{M_2|y\} \nonumber
\\
&&\quad =  \bigl[
\mathrm{P}\{y|M_1\}/\mathrm{P}\{y|M_2\}\bigr] \bigl[
\mathrm{P}\{M_1\}/\mathrm{P}\{M_2\}\bigr]
\\
&&\quad= [Z_{M_1}/Z_{M_2}] \bigl[\mathrm{P}\{M_1\}/
\mathrm{P}\{M_2\}\bigr].
\nonumber
\end{eqnarray}
A much maligned feature of the marginal
likelihood in this context is its possible sensitivity to
the choice of the parameter priors, $\mathrm{P}\{\theta|M_1\}$ and
$\mathrm{P}\{\theta|M_2\}$, through $Z_{M_1}$ and $Z_{M_2}$.  When limited
information is available to inform (or justify) this choice,  the
resulting Bayes factor can appear almost arbitrary.  [On the other
hand, viewed as a quantitative implementation
of Ockham's Razor, the key role of prior precision may well serve as
strong justification for the use of
Bayesian model selection in the scientific context; cf. \citet{jef91}.]  In their
influential treatise on this topic
\citet{kas95} thus argue that some form of prior-sensitivity
analysis be conducted as a routine part of all Bayesian model choice experiments,
their default recommendation being the recomputation of the Bayes
factor under a doubling and halving of key hyperparameters.

If the original marginal likelihoods have been estimated under an amenable
simulation scheme, then, as \citet{cho10} point out for
the case of nested importance sampling, alternative Bayes
\mbox{factors} under (moderate) prior
rescalings may be easily recovered by appropriately reweighting the
existing draws
without the need to incur further (computationally
expensive) likelihood function calls; and, indeed, the RLR method was
conceived specifically to facilitate such computations
(though in the reweighting mixtures context; \citealt{gey92}; \citealt{gey94}).
Using the $\hat{Z}_{k}$ from biased sampling under our nominal prior for a given model, the
pseudo-mixture density,
$p(\theta)$, of Equation (\ref{mixtured}) now serves as an efficient
``proposal'' for pseudo-importance sampling of various other targets with mass
concentrated near that of the posterior.
In particular, for the alternative
marginal likelihood, $\hat{Z}_\mathrm{alt}$, under some alternative
prior density, $\pi_\mathrm{alt}(\theta)$, we have
%
\begin{equation}
\label{priorsenseeqn} \hat{Z}_\mathrm{alt} = \sum_{i=1}^n
L(y|\theta_i) \pi_\mathrm{alt}(\theta_i)/p(
\theta_i)/n.
\end{equation}
The stability of this importance sample reweighting procedure may be monitored via
the effective sample size, $\mathrm{ESS} =
n/[1+\operatorname{var}_p\{\pi_\mathrm{alt}(\theta)/p(\theta)\}]$, following
\citet{kon94}, and its asymptotic variance estimated via recomputation
of Equation (\ref{priorsenseeqn}) \mbox{under} perturbations to the original $\hat{Z}_k$ drawn from the biased
sampling covariance matrix with bootstrap resampling of the pooled $\theta_i$.

For the general case of biased sampling from $w_j(\theta)$-weighted versions of a prior distribution,
$P_\theta$, not necessarily admitting a $\Lambda$-density, the equivalent formula takes the Radon--Nikodym derivative
of the alternative prior with respect to the original,
$\frac{dP_{\theta,\mathrm{alt}}}{dP_\theta}(\theta)$ (for $P_{\theta,\mathrm{alt}} \ll P_\theta$), such that
%
\begin{eqnarray}
\label{rnreweighting} \hat{Z}_\mathrm{alt} &=& \sum_{i=1}^n
L(y|\theta_i) \frac{dP_{\theta,\mathrm{alt}}}{dP_\theta}(\theta)
\nonumber
\\[-8pt]
\\[-8pt]
&&\phantom{\sum_{i=1}^n}{}\Big/ \Biggl[\sum
_{j=1}^m n_j/n \times w_j(
\theta_i)\Biggr] \bigg/n.\nonumber
\end{eqnarray}
We demonstrate the utility of this approach to prior-sensitivity
analysis in our finite and infinite mixture modeling of the well-known galaxy data set
in Section~\ref{mixtures}---and we refer the
interested reader to our other recent astronomical application concerning a
semiparameteric mixed effects model presented in \citet{cam13}.  Though
both these examples are based on the Dirichlet process prior, one can
\mbox{envisage} application of the same technique to investigate
prior sensitivity in many other problems of applied statistics---for example, Gaussian or
Ornstein--Uhlenbeck process modeling of astronomical time series (\citep{bre09}; \citep{bai12}).

\subsection{Designing and Sampling the Bridging Sequence}\label{fisher}

Although the recursive update scheme of biased sampling provides a powerful technique for
estimating the marginal likelihood given i.i.d. draws from a prespecified
sequence of $w_j(\theta)$-weighted distributions, the design of this
bridging sequence and the choice of an algorithm to sample from it are left
to the user.  While it is possible from theoretical principles to identify the optimal choice of
$w_j(\theta)$ with respect to the asymptotic variance under perfect
sampling for a limited range of problems---for example, \citet{gil88}
show the optimality of $w_1(\theta)=|L-Z|$ (requiring $Z$ known!) for the one sample case
with $F=P_L$ (in our marginal likelihood notation)---the design
problem cannot easily be solved in general. Moreover, even where a
theoretically optimal sequence can be identified, it will not
necessarily be computationally feasible to sample from such a
sequence.  Of more practical value therefore are heuristic guides for the
pragmatic choice of $w_j(\theta)$: strategies that will in a wide
variety of applied problems produce
adequate bridging sequences to ensure
manageable uncertainty in the output $\hat{Z}$ while remaining
accessible to existing posterior sampling techniques.  This topic
in various guises is the focus for the remainder of this paper, including our
numerical examples.

Perhaps the most natural family of bridging sequence for use on the
recursive pathway is that of the power posteriors method
[Equation (\ref{pp}); \citealt{lar06}; \citealt{fri08}]: this being both the
favored approach for past DoS-based applications
(\citep{hab12}; \citep{tan12})---where the
parameter, $t$, has a physical interpretation as the inverse
system temperature---and in Geyer's formulation of RLR---where
this particular sampling strategy ties in neatly with his parallel tempering MCMC
algorithm (MC$^3$; \citealt{gey92b}).  And, indeed, in
Section~\ref{tivis} below we will describe yet another conceptual connection between
these two methods, providing a heuristic justification for the borrowing
of thermodynamic integration strategies to this end.  Importantly, simulation from the power posterior at
an arbitrary $t_j$ is typically no more difficult than simulation from
the full posterior ($t_m=1$), the required modifications to a standard MCMC
and/or Gibbs sampling code being often quite trivial (e.g.,
\citealt{cam13}).  With biased sampling devised for i.i.d. draws, though,
it is important to thin the resulting chains (\citep{tan12}) so as not to bias the
corresponding asymptotic covariance estimates.  Experience has shown
that prior-focused temperature schedules, such as
$t=\{0,1/(m-1),2/(m-1),\ldots,1\}^c$ with $c\sim3$--5, tend to work well
for thermodynamic integration (\citep{fri08}), and we confirm this also
for biased sampling of our banana-shaped likelihood case study in
Section~\ref{banana}. [Likewise for tempering from a normalized auxiliary
density, $h(\theta)$,
closer in Kullback--Leibler divergence to the posterior than the prior;
\citet{lef10} and see our Section~\ref{tiauxiliary}.]

Another effective choice of bridging sequence for biased
sampling, which we demonstrate in our galaxy data set case study
of Section~\ref{mixtures}, is that of partial data posteriors (cf.
\citealt{cho02}): that is, $L(y^{(r_j)}|\theta)\pi(\theta)$ where $y^{(r_j)}$
represents a subset of $r_j$ elements of the full data set with $r_1=0$
the prior and $r_m=n_\mathrm{tot}$ the full posterior.  For i.i.d. $y$,
with an expected contribution of $r_j$ times the unit Fisher
information, the ``volume'' of highest posterior mass should shrink as
roughly $\sqrt{r_j}$, suggesting an automatic choice of roughly $r_j = \lfloor
n_\mathrm{tot}\times\{0,1/(m-1),2/(m-1),\ldots,1\}^c \rfloor$ with $c=2$ for this method.  (In
practice though, the first nonzero $r_j$ may well be limited by
sampling/identifiability constraints on the model; for our mixture model, for
instance, we must specify $r_2=k$, the number of mixture components.)

Finally, as observed by \citet{hab12}, the con\-strained-likelihood bridging
sequence of nested sampling can also be represented within the DoS
framework via $w_j(e) = I(e<e_{j})$ with $e_j < e_{j-1}$, although in
practice (as we explore in Section~\ref{banana}) the non-i.i.d. nature of
the resulting draws (with each draw from $e_{j-1}$ influencing the
placement of the next $e_{j}$ and its successors) violates the
assumptions of the biased sampling paradigm and ultimately limits the utility
of this approach by biasing its asymptotic covariance estimate.  In
fact, this issue more generally remains an open problem for recursive
marginal likelihood estimation theory: how can we best design effective
strategies for \textit{adaptively} choosing our bridging sequence, and
how can such modifications  to the biased sampling paradigm be accounted
for theoretically?  Given the effectiveness of empirical process
theory for characterizing the asymptotics of Vardi's biased sampling,
it seems likely that a solution to the above will require extensive
work in this area (with a focus on the impact of long-range
dependencies).  A similar problem arises in describing the asymptotics
of adaptive multiple importance sampling (\citep{cor12}), which without
its adaptive behavior could be considered a version of biased
sampling with known $W_j$; \citet{mar12} were recently able to provide
a consistency proof for a modified version of this algorithm, but with
a CLT remaining elusive.

\section{Thermodynamic Integration via Importance Sampling}\label{tivis}

Inspired by the recursive pathway of biased sampling, RLR and the DoS, we
present here yet another such strategy for marginal likelihood
estimation, which we name ``thermodynamic integration via importance
sampling'' (TIVIS).  Although quite novel at face value, it is easily
shown to be  a direct transformation of the recursive update methodology; yet by effectively
recasting this as a thermodynamic integration procedure we attain
insight into the relationship between its error budget and bridging sequence.  Specifically, the error in the estimation of
each $Z_{k}$ may be thought of as dependent on both the $J$-divergence
(\citep{lef10}) between it and the
remainder of the ensemble (via the thermodynamic identity) and on the accuracy of our estimates for
those other $Z_{j}$ ($j\neq k$).

To construct the TIVIS estimator, we once again assume the availability of pooled draws,
$\{\{\theta_i\}_j\dvtx i=1,\ldots,n_j; j=1,\ldots,m\}$, from a sequence
of bridging densities, $f_{j}(\theta) =
q_{j}(\theta)/Z_{j}$ ($j=1,\ldots,m$), with each
$g_{k}(\theta)$ exactly known.  Moreover, we suppose that $j=1$ indexes a
normalized reference/auxiliary, $\pi(\theta)$ or $h(\theta)$, such
that $Z_1=1$ is known, but with the remaining $Z_{k}$ typically unknown.  Despite our subsequent use of the thermodynamic
identity, however, we do not necessarily
require here that the bridging densities follow the geometric
path between these two extremes.  Now, rather than seek each
$\hat{Z}_{k}$ via direct importance sampling from $p(\theta)$ as
per the RLR, the TIVIS method is to instead seek each
normalization constant via thermodynamic integration from its preceding
density in the ensemble, $q_{k-1}(\theta)$, using the identity,
%
\begin{equation}
\label{fulltivis} \log Z_{k} = \int_0^1
E_{\pi_t^k} \bigl\{ \log \bigl( q_{k}(\theta)/f_{k-1}(
\theta) \bigr) \bigr\} \,dt,
\end{equation}
where $\pi_t^k(\theta) \propto
[f_{k}(\theta)]^t[f_{k-1}(\theta)]^{1-t} \propto
[g_{k}(\theta)]^t\times\break  [g_{k-1}(\theta)]^{1-t}$.  For existence
of the log-ratio in Equation (\ref{fulltivis}) we must impose
the strict condition (\textit{not} necessary for ordinary RLR) that all $f_{k}(\theta)$
share matching supports.  Pseudo-importance
sampling from $p(\theta)$---that is, importance sample reweighting of the
drawn $\{\theta_i\}_j$---allows construction of the appropriate (but
unnormalized) weighting function,
\[
u(\theta,t) = \bigl[g_{k}(\theta)\bigr]^t
\bigl[g_{k-1}(\theta)\bigr]^{1-t}/p(\theta),
\]
which in substitution to Equation (\ref{fulltivis}) yields the TIVIS
estimator,
%
\begin{eqnarray}
\label{tivismain}
&&\log (\hat{Z}_{k} / \hat{Z}_{k-1})\nonumber
\\
&&\quad= \int
_0^1 \Biggl[\sum_{i=1}^n
\log \bigl( g_{k}(\theta_i)/g_{k-1}(
\theta_i) \bigr) u(\theta_i,t)\Biggr]
\\
&&\phantom{\quad= \int
_0^1}{}\Big/ \Biggl[\sum
_{i=1}^n u(\theta_i,t)\Biggr]
\,dt.\nonumber
\end{eqnarray}
In computational terms, numerical solution of\break  the one-dimensional integral in the above may
be\break
achieved to arbitrary accuracy by simply evaluating the integrand at
sufficiently many $t_j$ on
the unit interval, followed by summation with
Simpson's rule.  If the sequence of bridging densities is
well chosen (and suitably ordered), the $J$-divergence between each
$f_{k}(\theta)$ and $f_{k-1}(\theta)$ pairing should be far less than that
between prior and posterior, such that a na\"ive regular spacing of the $t_j$
will suffice.

\begin{figure*}

\includegraphics{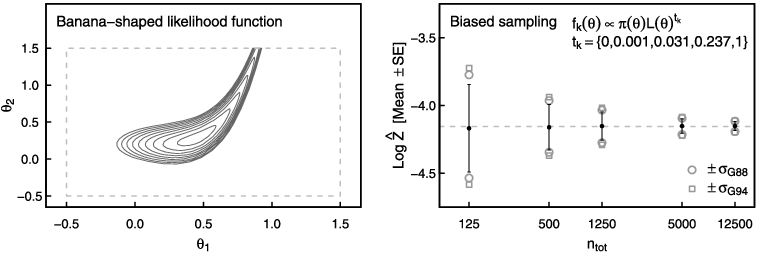}

\caption{The banana-shaped likelihood
function of our first case study [Equation (\protect\ref{bananafn}) of Section~\protect\ref{banana}] illustrated graphically as a
logarithmically-spaced contour plot on the domain of our Uniform
prior, $[-0.5,1.5]\times [-0.5,1.5]$ (left-hand panel).  Convergence of the
biased sampling estimator for the corresponding marginal likelihood
under MC$^3$ sampling of the power posterior (at five prespecified
temperatures) as a function of the total sample size is shown in the
right-hand panel.  The marked points and error bars on this figure indicate
respectively the
recovered mean and standard error (SE) in $\log \hat{Z}$ for 100 trials at
each $n_\mathrm{tot}$.  The dashed, light grey line indicates the \textup{``}exact\textup{''} $\log
Z$ for this example derived via brute-force quadrature, and the
open, light grey symbols indicate the mean \textup{``}per simulation\textup{''} estimate
of the SE from the asymptotic covariance matrix formulae of
Gill, Vardi and Wellner (\citeyear{gil88}) and Geyer (\citeyear{gey94}) alternately.}
\label{bananafig}
\end{figure*}

To show the equivalence between this estimator and that of the
recursive update scheme defined by Equation (\ref{Zupdates}), we
simply observe that the derivative of the denominator in Equation
(\ref{tivismain}) equals the numerator and, thus, by analogy to $\int_0^1
s^\prime(x)/s(x)\,dx = \log s(1) - \log s(0)$, we have
\begin{eqnarray*}
\log ( \hat{Z}_{k}/ \hat{Z}_{k-1}) &=& \log \Biggl[\sum
_{i=1}^n g_{k}(
\theta_i) /p(\theta_i)\Biggr]
\\
&&{}- \log \Biggl[\sum
_{i=1}^n g_{k-1}(
\theta_i) /p(\theta_i)\Biggr],
\end{eqnarray*}
and, thus,
\[
\log \hat{Z}_{k} = \log \Biggl[\sum_{i=1}^n
g_{k}(\theta_i) /p(\theta_i)/n\Biggr].
\]
In the following two case studies we further explore by numerical
example various issues concerning the design of the bridging sequence
[with particular reference to the efficiency in $L(y|\theta)$
calls; Section~\ref{banana}], and we highlight the utility of the \textit{normalized}
bridging sequence for prior-sensitivity analysis (Section~\ref{mixtures}).

\section{Case Study: Banana-Shaped LIkelihood~Function}\label{banana}

For our first case study we consider a (``mock,'' that is, data independent)
banana-shaped likelihood function, defined in two dimensions ($\theta=\{\theta_1,\theta_2\}$) as
%
\begin{eqnarray}
\label{bananafn} L(\theta) &=& \exp \bigl( -\bigl(10\times (0.45-
\theta_1)\bigr)^2/4
\nonumber
\\[-8pt]
\\[-8pt]
&&\phantom{\exp \bigl(\ \,\,}{}-\bigl(20\times \bigl(
\theta_2/2-\theta_1^4\bigr)
\bigr)^2\bigr) ,\nonumber
\end{eqnarray}
with a Uniform prior density of $\pi(\theta)=1/4$ on the
rectangular domain,
$[-0.5,1.5]\times [-0.5,1.5]$.  A~simple illustration of this
likelihood function as a
logarithmical\-ly-spaced contour plot is presented in the
left-hand panel of Figure~\ref{bananafig}.  Brute-force numerical integration via
quadrature returns the ``exact'' solution, $Z=0.01569[6]$ (or $\log Z
= -4.154[3]$).

\begin{figure*}[b]

\includegraphics{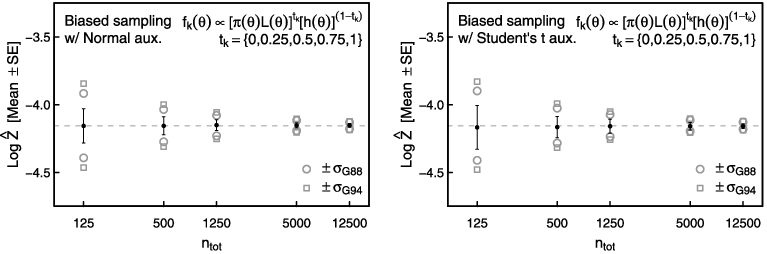}

\caption{Convergence of the
biased sampling estimator for the marginal likelihood of our
banana-shaped likelihood function
under MC$^{3}$ sampling
(at five prespecified
temperatures) on the geometric path between a \textup{``}data-driven\textup{''}
reference/auxiliary density, $h(\theta)$, and the posterior, shown as a
function of the total sample size. The adopted $h(\theta)$ takes a two-dimensional
Student\textup{'}s $t$ form in the left-hand panel and a Normal form in the right-hand,
with its controlling parameters ($\mu_\mathrm{mode}$ and
$\Sigma_\mathrm{mode}$)  in each case set to the location and
curvature of the posterior mode.  A marked reduction in standard
error (at fixed $n_\mathrm{tot}$) with respect to that of the na\"ive (power
posteriors) path, that is, $h(\theta)=\pi(\theta)$, is evident from comparison with Figure~\protect\ref{bananafig}.}
\label{tibanana}
\end{figure*}

As a benchmark of the method we first apply the biased sampling
estimator to draws from a sequence of bridging densities
following the standard
power posteriors path. Though even a cursory inspection of the likelihood
function for this simple case study is sufficient to confirm its
unimodality and to motivate a family of suitable proposal densities for straightforward importance
sampling of $\pi_t(\theta) \propto \pi(\theta)L(\theta)^t$, for
demonstrative
purposes we have chosen to implement
an MC$^3$ (\citep{gey92b}) approach here instead, the
latter being ultimately amenable to more complex
posteriors than the former.  Following standard practice for
thermodynamic integration---as per our motivation from Sections~\ref{fisher} and~\ref{tivis} above---we adopt a prespecified
tempering schedule spaced geometrically as
$t=\{0,1/(m-1),2/(m-1),\ldots,1\}^c$ with $c=5$ and $m=5$.  To
illustrate the $1/\sqrt{n}$ convergence of biased sampling, we run this
procedure 100 times at each of five total sample sizes
($n_\mathrm{tot}=\{125,500,1250,5000,12\mbox{,}500\}$; distributed equally
across all five temperatures) thinned at a rate of 0.25
from their parent MC$^3$ chains. The
resulting mean and standard error (SE) at each $n_\mathrm{tot}$ are marked in the right-hand panel
of Figure~\ref{bananafig}.

Overlaid are (the means of) the
corresponding  ``per simulation'' estimates of this
standard error computed from the rival asymptotic covariance matrix
forms of  \citet{gil88}/\citet{kon03} and \citet{gey94}: the former
being originally derived
from the empirical process CLT applicable to biased sampling and the
latter from  maximum likelihood theory using the Hessian of the quasi-likelihood
function for reverse logistic regression.  As noted in Section~\ref{bs},
\citet{kon03} have previously discussed the inadequacy of
Geyer's covariance estimator---though for the present design the
difference is negligible.  It is worth noting that both estimates are
a little conservative at low $n_\mathrm{tot}$ but give an excellent
agreement with the repeat simulation SE by $n_\mathrm{tot}=1250$.

With this power posteriors version of biased sampling as benchmark, we now consider the merits of two alternative
schemes for defining, and sampling from, the required sequence of bridging
densities, $f_{k}(\theta)$, in Sections~\ref{tiauxiliary} and
\ref{nestedcomp} below.

\subsection{Thermodynamic Integration from a~Reference/Auxiliary Density}\label{tiauxiliary}

As highlighted by \citet{lef10}, the error budget of
thermodynamic integration over the geometric path depends to first-order upon the
$J$-divergence between the reference/auxiliary density, $h(\theta)$,
and the target, $\pi(\theta|y)$.
Thus, it will generally be more efficient to set a ``data-driven''
$h(\theta)$---such as may be recovered from the position and local
curvature of the posterior mode---than to integrate ``na\"ively'' from the prior, that is,
$h(\theta)=\pi(\theta)$.  Here we demonstrate the corresponding
improvement to the performance of the biased sampling estimator
resulting from the choices, $h(\theta)\sim
\mathcal{N}_\mathrm{Trunc.}(\mu_\mathrm{mode},\Sigma_\mathrm{mode}^{-1})$
and $h(\theta)\sim
\mathcal{T}_\mathrm{Trunc.}(\mu_\mathrm{mode},\Sigma_\mathrm{mode}^{-1})$.
Here
$\mathcal{N}_\mathrm{Trunc.}$ and $\mathcal{T}_\mathrm{Trunc.}$ denote
the two-dimensional Normal and Student's
$t$ ($\nu=1$) distributions (truncated to our prior support), respectively, while $\mu_\mathrm{mode}$
denotes the posterior mode and
$\Sigma_\mathrm{mode}$ its local curvature (recovered here
analytically, but estimable at minimal cost in many Bayesian analysis
problems via standard numerical methods).  As before, we apply MC$^3$
to explore the tempered posterior and repeat both experiments 100
times at each of our five $n_\mathrm{tot}$.  In contrast to the power posteriors
case, we adopt here a regular temperature
grid, $t=\{0,0.25,0.5,0.75,1\}$, to allow for the imposed/intended
similarity between $\pi(\theta|y)$ and $h(\theta)$. Our
results are presented in Figure~\ref{tibanana} and discussed below.

\begin{figure*}

\includegraphics{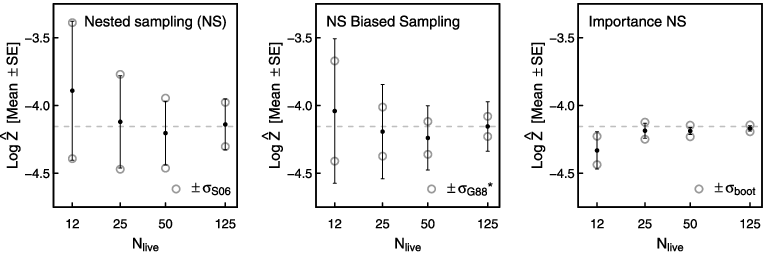}

\caption{The performance of nested sampling (left-hand panel) as a
marginal likelihood estimator for our banana-shaped
pseudo-likelihood function, run under the
ellipse-based strategy for exploring the sequence of
constrained-likelihood densities
proposed by Mukherjee, Parkinson and Liddle (\citeyear{muk06}); compared with that of biased sampling
(middle panel) and \textup{``}importance nested sampling\textup{''} (right-hand panel) with the same
bridging sequence.  The
first two schemes converge to the true $\log Z$ at a similar rate in
$N_\mathrm{live}$, while the third is faster since it harnesses the
information content of draws otherwise discarded from nested sampling
in its constrained-likelihood search.}
\label{nestedbanana}
\end{figure*}

As expected from both theoretical considerations (\citep{gel98}; \citep{lef10})
and reports of practical experience with other marginal likelihood
estimators (\citep{fan12}),  use of a ``data-driven'' auxiliary in this
example has
indeed reduced markedly the standard error of the biased sampling scheme
(at fixed $n_\mathrm{tot}$) with respect to that of the na\"ive (power
posteriors) path, that is, $h(\theta)=\pi(\theta)$.  In this instance
the (thinner-tailed) Normal auxiliary  has outperformed the
(fatter-tailed) Student's $t$ (with one d.o.f.); however, although
this result is
again consistent with theoretical expectations---as a quick computation using the ``exact''
$\log Z$ confirms
$J[\mathcal{N}(\mu_\mathrm{mode},\Sigma_\mathrm{mode}^{-1}),h(\theta)]
\ll
J[\mathcal{T}(\mu_\mathrm{mode},\Sigma_\mathrm{mode}^{-1}),  h(\theta)]$---it
should be remembered that the optimal choice of auxiliary from within a standard
parametric family  depends
on the likelihood function itself, and so will vary from problem to problem.
Moreover, without knowledge of the desired $Z$ it is not possible
to optimize $h(\theta)$ {a priori}; and even a crude
estimator of the $J$-divergence run with, for example, the Laplace approximation
to the marginal likelihood will nevertheless add numerous extra likelihood
evaluations to the computational budget.  Although ``fatter-tailed''
than a typical likelihood function, the Student's $t$ may well prove a
superior choice for some multimodel posterior problems in practice by better facilitating mixing during the MC$^3$ sampling stage.

\subsection{Ellipse/Ellipsoid-Based Nested Sampling}\label{nestedcomp}

Recalling the connections between the DoS derivation of the recursive
pathway and the nested sampling algorithm described in Section~\ref{dossection}, it is of some
interest to
compare directly the performance of these \mbox{rival} techniques.  The present
case study with its Uniform prior density is in fact well suited to this purpose since
in the field of cosmological model selection, where nested
sampling has been most extensively used of late (\citep{muk06}; \citep{fer08}), it is standard practice to
adopt separable priors from which a Uniform sample space
may be easily constructed under the
quantile function transformation, which, for the discussion below, we assume has been done such
that $\pi(\theta)$ may be taken as strictly Uniform on $[0,1]^N$ (in
the transformed coordinate space).  Given
these conditions, \citet{muk06} outline a crude-but-effective scheme
for exploring the constrained-likelihood shells of nested sampling, in
which the new ``live'' particle for each update must be drawn with
density
proportional to $\pi(\theta)I(L(\theta)>L(\theta_{i-1}))$.

Under the
\citet{muk06} scheme, to draw the required $\theta_i$, one simply identifies the minimum bounding
ellipse [or with $D(\theta)>2$, the minimum bounding
\textit{ellipsoid}] for the present set of ``live'' particles, expands
this ellipse by a small factor $\sim$1.5--2 with the aim of enclosing the full
support of $I(L(\theta)>L(\theta_{i-1}))$, and then draws
randomly from its interior until a valid $\{\theta_i,L(\theta_i)\}$ is discovered.
Supposing the elliptical sampling window
thus defined has been enlarged sufficiently to fully enclose the desired
likelihood surface [which it must do to ensure unbiased sampling of
$\{\theta_i,L(\theta_i)\}$, although we can rarely be \textit{sure} that it has],
it remains unlikely to match its shape exactly, leading to
an overhead of $n_\mathrm{oh}$ discarded draws, $\{\theta_i^{(j)}\dvtx L(\theta_i^{(j)}) <
L(\theta_{i-1}), j=1,\ldots,n_\mathrm{oh}\}$. At each $\theta_i$
the incurred $n_\mathrm{oh}$ may be thought of as a single realization of the
negative binomial distribution with $p$ equal to the fraction of the
bounded ellipse for which $L(\theta) < L(\theta_{i-1})$, hence, $E(n_\mathrm{oh})=1/p-1$.  The magnitude of
this overhead can in general be expected to scale with the geometric
volume of the parameter space, potentially limiting the utility of this otherwise dimensionally-insensitive
Monte Carlo-based estimator.  However, where applicable, the
\citet{muk06} scheme may nevertheless prove more efficient than the alternative of
constrained-MCMC-sampling to find the new $\theta_i$ (cf. \citealt{fri12})
in which one must discard at least $\sim$10--20 burn-in moves [each
with a necessary $L(\theta)$ call] per step
to
achieve approximate stationarity.

Applying the ellipse-based approach to nested sampling of the
banana-shaped likelihood
function of Equation (\ref{bananafn}) with $N_\mathrm{live}=\{12,25,50,125\}$ live
particles evolved over $10\times N_\mathrm{live}$ steps in each
case [and a small extrapolation of the mean $L_\mathrm{live}$ times
$\exp(-10)$ at the final step; cf. \citealt{ski06}], we recover a convergence to
the true $\log Z$ as shown in the left-hand panel of Figure~\ref{nestedbanana}.
Important to note is that with the ellipse scale
factor of 1.5 used here the result is an overhead of
$n_\mathrm{oh}\approx 2.3$ likelihood calls per accepted $\theta_i$,
such that nested sampling at $N_\mathrm{live}=125$ corresponds to
$n_\mathrm{tot}\approx 2875$ in the previous examples.  An overhead of
this magnitude should be a
concern for ``real world'' applications of nested sampling in which the likelihood function
may be genuinely expensive to evaluate; indeed, for modern cosmological
simulations MCMC exploration of the $D(\theta)\lesssim 12$ posterior is effectively a
super-computer-only exercise due solely to the cost of solving for
$L(y|\theta)$. [At this point the skeptical reader might object that the distinctly nonelliptical $L(\theta)$ considered
in this example be considered
a particularly unfair case for testing the \citet{muk06} method,
but such
banana-shaped likelihoods are in fact quite common in higher-order
cosmological models; see, for instance, \citet{dav07}.]  We therefore suggest
that one might improve upon the efficiency of
ellipse-based nested sampling by co-opting its bridging sequence into
the biased sampling framework in some manner.

As \citet{hab12} has pointed out, the nested sampling pathway can be
accommodated roughly within the DoS (and hence biased sampling)
framework, for example, by treating the accepted $\theta_i$ (pooled with the
surviving $N_\mathrm{live}$ live particles) as drawn from the series
of weighted distributions, $w_j(\theta)\,dF(\theta) = I(L(\theta)>L_j)\,dF(\theta)$. However, with
each $w_j(\theta)$ ($j > 1$) now dependent on past draws---and hence
the $\{\theta_i\}_j$ no longer i.i.d.---although we can apply the
recursive update scheme of Equation (\ref{Zupdates}) to normalize the
bridging sequence and then importance sample reweight to $Z$, the
biased sampling CLT no longer holds.  To demonstrate this, we apply the
above procedure to the draws from our previous nested sampling runs
and plot the mean and repeat simulation SE at each $N_\mathrm{live}$
in the middle panel of Figure~\ref{nestedbanana}.  While the
efficiency of this estimator is almost identical to that of ordinary
nested sampling, the ``na\"ive'' application of Gill et al.'s asymptotic
covariance matrix does not yield an SE estimate matching that of repeat
simulation.

A more interesting alternative is to observe that for the
ellipse-based nested sampling method (given uniform priors) the
normalization of each $f_k(\theta)$ is in fact easily computed from
the area/volume of the corresponding ellipse/ellipsoid.  That is, we
can simply pool our draws---\textit{including} the $\theta_i$ with
$L(\{\theta_i\}_j) < L_j$ otherwise discarded from nested
sampling---and apply the importance sample reweighting procedure of
Equation~(\ref{priorsenseeqn}) with
$\pi_\mathrm{alt}(\theta)=\pi(\theta)$ and
\[
p(\theta) = \sum_{k=1}^m
[n_k/n_\mathrm{tot}] \bigl[I\bigl(\theta \in
\mathit{Ell}[E_{\mathrm{live}(k)}]\bigr)/V_k\bigr]
\]
(with $V_k$ the volume of the $k$th ellipse and
$\mathit{Ell}[E_{\mathrm{live}(k)}]$ its interior).  Application of this
strategy---which we dub ``importance nested sampling'' (INS)---to the
pres\-ent example yields $\log Z$ estimates with much smaller repeat
simulation SE than either of the previous summations as shown in the
right-hand panel of Figure~\ref{nestedbanana}.  Bootstrap resampling of
the drawn $\{\theta_i, L_i\}$ gives a reasonable estimator of this SE,
though we note that INS does not appear to be unbiased in $\log Z$,
with a slight tendency toward underestimation at small
$n_\mathrm{tot}$.  Further computational experiments are now underway to
better quantify the advantages offered by this approach to
harnessing the information content of these otherwise discarded
draws in the ellipse-based nested sampling paradigm (presented in
\citealt{fer13}).

\section{Case Study: Normal Mixture Modeling of the Galaxy Data Set}\label{mixtures}

The well-known galaxy data set, first proposed as a test case
for kernel density estimation by \citet{roe90}, consists of precise
recession velocity measurements (in units of 1000 km\,s$^{-1}$) for 82
galaxies in the Corona Borealis region of the Northern
sky reported by \citet{pos86}.  The purpose of the original
astronomical study was to search---in light of a then recently discovered
void in the neighboring Bo\"otes field (\citep{kir81})---for further large-scale
inhomogeneities in the distribution of galaxies.  Given the well-defined
selection function of their survey, \citet{pos86} were easily
able to compute as a benchmark the
recession velocity density function expected under
the null hypothesis of a uniform distribution of galaxies throughout
space, and by visual comparison
of this density against a histogram of their
observed velocities the astronomers were able to establish strong
evidence against the null, thereby boosting support for the (now canonical)
hierarchical clustering model of cosmological mass assembly (\citep{gun72}).  However,
under the latter
hypothesis, as \citet{roe90}
insightfully observed, one can then ask
the more challenging statistical question of ``\textit{how many distinct
clustering components are in fact present in the recession
velocity data set?}''

Many authors have since attempted to answer this
question (posed for simplicity as a univariate\vadjust{\goodbreak} Normal mixture
modeling problem) as a means to demonstrate the utility of their
preferred marginal likelihood estimation or model space exploration strategy.
Notable such contributions to this end include the following: the infinite mixture model (Dirichlet process
prior) analyses of
\citet{esc95} and \citet{phi96}; Chib's exposition of marginal
likelihood estimation from Gibbs sampling output (\citep{chi95}); the
reversible jump MCMC approach of \citet{ric97}; and the label
switching studies of \citet{ste00} and \citet{jas05}.  The earliest of
these
efforts are well summarized
by \citet{ait01}, who highlights a marked dependence of the
inferred number of mixture components on the chosen priors.  For
this reason, as much as its historical significance, the galaxy
data set provides a most interesting case study with which to illustrate the
potential of prior-sensitivity analysis under the
recursive pathway.

The outline of our presentation is as follows.
In Section~\ref{statmod} we set forth the finite and infinite mixture
models to be examined here and in Section~\ref{gibbssampling} we
describe the MCMC strategies we use to explore their complete and
partial data posteriors.  In Section~\ref{hyperp} we discuss
various astronomical motivations for our default hyperprior choices
and, finally, in Section~\ref{priorsense} we present the results of
a biased sampling run on this problem with importance sample
reweighting-based transformations between alternative priors.

\subsection{Normal Mixture Model}\label{statmod}

\subsubsection{Finite mixture model}

Following
\citet{die94} and
\citet{lee08}, we write the $k$-component Normal mixture
model with component weights, $\phi$, in the
latent allocation variable form for data vector, $y$, and
(unobserved) allocation vector, $z$, such that\vspace*{-1pt}
\[
\pi(z_i=j)=\phi_j \quad\mbox{and}\quad
\pi(y_i|z_i=j) = f_\mathcal{N}(y_i|
\theta_j).
\]
Here $f_\mathcal{N}(\cdot|\theta_j)$ represents the one-dimensional Normal
density, which we will reference in mean--precision syntax as
$\mathcal{N}(\mu_j,\tau_j^{-1})$, that is,
$\theta_j=\{\mu_j,\tau_j\}$.

Given priors for the number of components in the mixture, the
distribution of weights at a given $k$ and the vector of
mean precisions---that is, $\pi(k)$, $\pi(\phi|k)$ and $\pi(\theta|\phi,k)$, respectively---the posterior for the number of
mixture components in the \textit{finite} mixture case may be
recovered by integration over $\{\phi,\theta\}$ at each $k$,\vspace*{-1pt}
\begin{eqnarray*}
\pi(k|y) &=& \pi(k)/Z
\\
&&{}\times \int_\Omega \pi(\phi|k)\pi(\theta|
\phi,k)L(y|\theta,\phi,k)\,d\phi \,d\theta
\\
&=& \pi(k)Z^{(k)}/Z.
\end{eqnarray*}
Here the likelihood,
$L(y|\theta,\phi,k)$, is given by a summation over the $n_\mathrm{tot}$ unobserved,
$z_i$, as\vspace*{-1pt}
%
\begin{equation}
\label{likes} L(y|\theta,\phi,k) = \prod_{i=1}^{n_\mathrm{tot}}
\sum_{j=1}^{k}\phi_j
f_\mathcal{N}(y_i|\theta_j).
\end{equation}
That is, for a $\pi(k)$ assigning
mass to only a small set of elements,
one approach to recovering $\pi(k|y)$ is simply to estimate the ``per
component'' marginal likelihood, $\hat{Z}^{(k)}$, at each of
these $k$ and then reweight by $\pi(k)$.  The full marginal likelihood of the
model can then of course be estimated from the sum, $\hat{Z}=\sum_k \hat{Z}^{(k)}$.
While this is indeed the strategy adopted here for exposition purposes, it is worth noting that such direct marginal likelihood estimation to
recover $\pi(k|y)$
for this model can in fact be entirely avoided via either the reversible jump
MCMC algorithm (\citep{ric97}) or Gibbs sampling over the infinite
mixture version described below.

\subsubsection{Infinite mixture model}

Rather than specify a maximum
number of mixture components  {a priori}, \citet{esc95} and
\citet{phi96} (among others) have advocated an infinite-dimensional
solution based on the
Dirichet process prior.  In particular, one may suppose the data to
have been drawn from
an infinite mixture of Normals with means,
variances and weights drawn as the realization, $Q$, of a Dirichlet process (DP), $\operatorname{DP}(M,G_0)$, on
$\mathbb{R}\times\mathbb{R}_+$, the characterization of the DP being
via a concentration index, $M$, and reference density, $G_0$, and with all
$Q$ being both normalized and strictly atomic.  For small
$M$ ($\lesssim10$) the tendency is for these $Q$ to be
dominated by only a few (mixture) components, while for large $M$ the
number of significant components inevitably increases, with the
typical $Q$
thereby becoming closer (in the metric of weak convergence)
to $G_0$.  The likelihood of i.i.d. $y$ for a given $Q$ requires (in theory) an infinite
sum over the contribution from each of its components,
\[
L(y|Q) = \prod_{i=1}^{n_\mathrm{tot}} \sum
_{j=1}^\infty \phi_j f_\mathcal{N}(y_i|
\theta_j),
\]
where each $\phi_j$ represents the limiting fraction of points in the
realization assigned to a particular $\theta_j$.\vadjust{\goodbreak}  (In practice,
however, this summation can generally be truncated with negligible
loss of accuracy after \mbox{accounting} for the contributions of only the most dominant
components.)  Computation of
the marginal likelihood for the above model is thus nominally by
integration over the infinite-dimensional
space of
$Q$.  In particular, if we suppose a hyperprior density for the
hyperparameters, $\psi$, of the DP (i.e., for $M$ and the controlling
parameters of $G_0$), we have $Z = \int_{\Omega(\psi)}\int_{\Omega(Q)} L(y|Q)
\{dP_{Q|\psi}(Q)\} \pi(\psi) \,d\psi$.

As per the finite mixture case, we can simplify our posterior
exploration and relevant
computations by introducing latent variables, $z$ and $\theta$, for
allocation of the $y$ and the corresponding mean-precision vectors of
the parent components in $Q$.  In this version the likelihood takes
the form
\[
L\bigl(y|\{z,\theta\}\bigr) = \prod_{i=1}^{n_\mathrm{tot}}
f_\mathcal{N}(y_i|\theta_{z_i}),
\]
and the marginal likelihood becomes
%
\begin{eqnarray}
\label{vv1} Z &=& \int_{\Omega(\psi)}\int_{\Omega(Q)}
\int_{\Omega(\{z,\theta\})}L\bigl(y|\{z,\theta\}\bigr)
\nonumber
\\
&&\phantom{\int_{\Omega(\psi)}\int_{\Omega(Q)}
\int_{\Omega(\{z,\theta\})}}{}\cdot\bigl\{dP_{\{z,\theta\}|Q}
\bigl(\{z,\theta\}\bigr)\bigr\}
\\
&&\phantom{\int_{\Omega(\psi)}\int_{\Omega(Q)}
\int_{\Omega(\{z,\theta\})}}{}\cdot\bigl\{dP_{Q|\psi}(Q)\bigr\} \pi(\psi) \,d
\psi.\nonumber
\end{eqnarray}
Importantly, existing Gibbs sampling methods for the DP allow
for collapsed sampling from the posterior for $\{z,\theta,\psi\}$ and
Equation (\ref{vv1}) can be reduced to $
\int_{\Omega(\{z,\theta,\psi\})}L(y|\{z,\theta\})
\{dP_{\{z,\theta,\psi\}}(\{z,\theta,\psi\})\}$.  In one further twist,
however, we note that since the reduced expression is degenerate across
component labelings, it is in fact more computationally efficient to
estimate $Z$ from
%
\begin{eqnarray}
\label{likeq} &&\int_{\Omega(\{z,\theta,\psi\})} \int_{\Omega(\hat{Q})}L(y|
\hat{Q})
\nonumber
\\
&&\phantom{\int_{\Omega(\{z,\theta,\psi\})} \int_{\Omega(\hat{Q})}}{}\cdot\{dP_{\hat{Q}|\{z,\theta,\psi\}}\}
\\
&&\phantom{\int_{\Omega(\{z,\theta,\psi\})} \int_{\Omega(\hat{Q})}}{}\cdot\bigl\{dP_{\{z,\theta,\psi\}}\bigl(\{z,\theta,\psi\}
\bigr)\bigr\},\nonumber
\end{eqnarray}
where $P_{\hat{Q}|\{z,\theta,\psi\}}$ takes a particularly simple
analytic form by the nature of the DP (cf. \citealt{esc95}).

Finally, it is important to note that since each realization of the DP has always an infinite number of components
with probability one (though  usually only a few\vadjust{\goodbreak} with significant mass), the usual interpretation for the posterior,
$\pi(k|y)$, in this context is the posterior distribution of the
number of unique label assignments \textit{among the observed
data set} (i.e., the dimension of $\theta$ in $\{z,\theta\}$).  However, although pragmatically useful for such modeling
problems as that exhibited by the galaxy data set, as \citet{mil13} note, this estimate is not consistent.

\subsection{MC Exploration of the Mixture Model Posterior}\label{gibbssampling}

\subsubsection{Finite mixture model}\label{fmm}

Exploration of the posterior for $\{\theta,\phi\}$ at fixed $k$ in
this finite mixture model can be accomplished rather efficiently
(modulo the well-known problem of mixing \textit{between} modes; cf.
\citealt{nea99}) via Gibbs sampling given conjugate prior choices, as
explained in detail by \citet{ric97}.  To this end, we suppose
\begin{eqnarray*}
\mu_j &\sim& \mathcal{N}\bigl(\kappa,\xi^{-1}\bigr), \quad
\tau_j \sim \Gamma(\alpha,\beta),\quad\mbox{and}
\\
 \beta &\sim&
\Gamma(\beta_1,\beta_2),
\end{eqnarray*}
where $\Gamma(a,b)$ represents the Gamma distribution with rate $a$
and shape $b$.  To simulate from the resulting posterior, we use the
purpose-built code provided by \mbox{\texttt{BMMmodel}} and \texttt{JAGSrun}
in the \texttt{BayesMix} package (\citep{gru10}) for
\texttt{R}.  No modifications to this code are necessary for sampling
the partial data posterior, and both the partial and full data
likelihoods given partial likelihood draws (at fixed $k$) may be
recovered with Equation (\ref{likes}).  The range of $k$ for which we
compute marginal likelihoods is here limited by the range of a truncated Poisson prior on $k$.

\subsubsection{Infinite mixture model}

As noted earlier, exploration of the infinite mixture model posterior can also be facilitated through Gibbs sampling with the appropriate choice of
priors (\citep{esc95}); and although contemporary codes typically use
the (more efficient) alternative algorithm of \citet{nea00}, the prior forms dictated by
the conjugacy necessary for Gibbs sampling remain the default.  Hence, to this end, we suppose
a fixed concentration index of $M=1$ and
a Normal-Gamma reference density,
\[
G_0\dvtx \tau_j \sim \Gamma(s/2,S/2), \quad
\mu_j|\tau_j \sim \mathcal{N}(m,\tau_j h),
\]
assigning hyperpriors of $h \sim \Gamma(h_1/2,h_2/2)$ and $1/S \sim
\Gamma(\nu_1/2,\nu_2/2)$. Here we use the \texttt{DPdensity} function
in the
\texttt{DPpackage} (\citep{jar11}) for \texttt{R} to explore this
posterior.  While no modifications to this code are required\vadjust{\goodbreak} for sampling the
partial likelihood posteriors, the computation of full
data likelihoods given the partial likelihood posterior requires that
we sample a series of dummy components from the current posterior until
some appropriate truncation point, $k^\prime\dvtx  \sum_{j=1}^{k^\prime} \phi_j \approx 1$, before
applying (the $k^\prime$-truncated version of) Equation (\ref{likeq}).

\begin{figure*}

\includegraphics{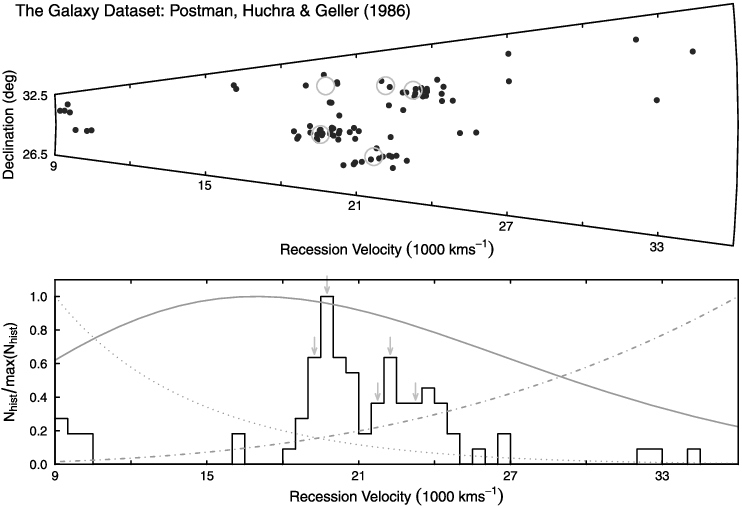}

\caption{Visualization of the galaxy data set, including
its Abell clusters and selection function.  The clustering of
galaxies in  recession velocity--declination space
is illustrated by way of the \textup{``}cone diagram\textup{''} shown in the top
panel and its projection to a recession velocity
histogram shown in the bottom panel.  The positions of five Abell clusters
targeted by the original survey are also highlighted here (open circles
and arrows in light grey), along with the survey\textup{'}s magnitude-dependent, volume-dependent and
net selection functions (shown as the dotted, dash-dotted
and solid curves, respectively, in the bottom panel).}
\label{postman}
\end{figure*}

\subsection{Astronomical Motivations for our Priors}\label{hyperp}

\subsubsection{Finite mixture model}

As noted earlier, by considering the
well-defined selection function of their observational campaign, the
authors of the original astronomical study were able to construct the
expected probability density function of
recession velocities for their survey under the null hypothesis
of
a
uniform distribution of galaxies throughout space.  In
particular, \citet{pos86} recognized that the strict \textit{apparent}
magnitude limit of their spectroscopic targeting strategy ($m_r <
15.7$ mag) would act as a luminosity (or \textit{absolute} magnitude) limit
evolving with recession velocity (distance) according to
\[
M_{r,\mathrm{lim}}(v) \approx m_r - 5\log_{10}(v) - 30,
\]
where we have assumed units of 1000 km\,s$^{-1}$ for $v$ and a ``Hubble
constant'' of $H_0 = 100$ km\,s$^{-1}$\,Mpc$^{-1}$.  To estimate the form
of the resulting selection function, $S_\mathrm{mag}(v)$,
\citet{pos86} considered how the relative number of galaxies per unit
volume  brighter than this limit would vary with
distance given the absolute magnitude distribution function, $F_\mathrm{mag}(\cdot)$, for
galaxies in the local Universe, that is, $S_\mathrm{mag}(v) \propto
1-F_\mathrm{mag}(M_{r,\mathrm{lim}}(v))$.   To approximate the latter,
the astronomers simply integrated over a previous
estimate of the local luminosity density parameterized as a Schechter
function (\citep{sch76}) with characteristic magnitude,
$M_r^\ast\approx-19.40-1.5$ mag,
and faint-end slope, $\alpha_r^\ast\approx-1.3$, such that
\[
f(M) \propto \bigl[10^{2/5(M_r^\ast-M)}\bigr]^{\alpha_r^\ast+1} \exp
\bigl[-10^{2/5(M_r^\ast-M)}\bigr]
\]
and
\[
S_\mathrm{mag}(v) \propto \int_{-\infty}^{M_{r,\mathrm{lim}}(v)}f(M)
\,dM.
\]

An interesting feature of magnitude-limited astronomical surveys is
that, although with increasing recession velocity this $S_\mathrm{mag}(v)$ selection
function restricts their sampling to the decreasing fraction of
galaxies above $M_{r,\mathrm{lim}}(v)$, the volume of the Universe probed by (the projection into
three-dimensional space of) their two-dimensional angular viewing window
is, in contrast, rapidly
increasing.  Hence, there exists an important additional selection
effect, $S_\mathrm{vol}(v)$, operating in competition with, and
initially dominating, that on
magnitude, and scaling
with (roughly) the third power of recession velocity such that
\[
S_\mathrm{vol}(v) \propto v^3.
\]

The product of these two effects therefore returns the net selection function of
the galaxy data set, which we illustrate (along with each effect in
isolation) in Figure~\ref{postman} (see also Figure~4b from
\citealt{pos86});  the point being that there do exist informative
astronomical considerations for choosing at least some of the
hyperparameters of our priors in this mixture modeling case study,
though past analyses have tended to ignore this context (contributing somewhat to the apparent ``failure'' of Bayesian mixture
modeling for this data set; \citealt{ait01}).  In
particular, the shape of the selection function in velocity space
suggests the form for our prior on the component means: a choice of
$\{\kappa=17,\xi=0.008\}$ gives a
reasonable match to the shape of $S_\mathrm{mag}(v)
S_\mathrm{vol}(v)$.  Perhaps surprisingly, as we will demonstrate
later, the choice of prior on the component means has a substantial
influence on the resulting $\pi(k|y)$; changing only these of our
hyperparameters to ``data-driven''  values chosen as $\{\kappa=\bar{y}\
(20.8),\xi=1/\operatorname{var}\{y\}\ (0.048)\}$
results in a drastic shift of the posterior.

Likewise, we can inform our prior choice for the number of components
in the mixture with regard to the original survey design, which
featured five separate observational windows placed so as to cover five previously identified galaxy clusters from the
Abell catalog. (The positions of these clusters in bivariate recession
velocity--declination space, and its projection to univariate velocity
space, are also marked on Figure~\ref{postman} for reference.)  Hence, we select a
mode of $\lambda=5$ for our truncated
Poisson prior for $\pi(k)$.  With the $k=1$ and
$k=2$ mixture models already well excluded by previous analyses, and
$k > 10$ a pragmatic upper bound for exploration given $\lambda
=5$, we therefore truncate our prior to the range $3 \le k \le 10$.  This
contrasts somewhat with the Uniform priors on $k\le10$ and $k\le30$ assumed by
\citet{roe97} and \citet{ric97}, respectively---though reweighting for
alternative $\pi(k)$ (on this support) is trivial in any case.

Only the
precisions of the Normal mixture components are not well constrained
from astronomical considerations---although we can at least be confident
that any large-scale clustering should occur above the scale of
individual galaxy clusters ($\sim$1 Mpc or $\Delta v \approx 0.1$) and
(unless the uniform space-filling hypothesis were correct) well below
the width of our selection function.  Thus, we simply adopt a
fixed shape hyperparameter of $\alpha=2$ for our Gamma prior on
the $\tau_j$ and allow the rate
hyperparameter to vary according to its Gamma hyperprior form
$\beta_1=1$ and $\beta_2=0.05$.  Our choice here is thus comparable to
that of \citet{ric97} who suppose
$\pi(\beta)\sim\Gamma(0.2,0.016)$---not $\Gamma(0.2,0.573)$ as
misquoted by \citet{ait01}---though we evidently place far less
prior weight
on exceedingly large precisions (small variances).

\begin{figure*}

\includegraphics{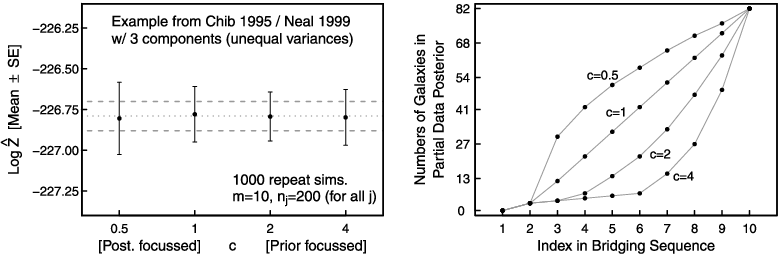}

\caption{Relationship between the standard error (SE) of $\log Z$ estimation and the choice of
a partial data posterior bridging sequence for biased sampling of the 3-component (unequal variance)
mixture model under the Chib (\citeyear{chi95})
priors.  The data points in the left-hand panel represent the mean
$\log \hat{Z}$ and the error bars its (single
run) SE, computed from 1000 repeat simulations with 200
draws from each of 10
steps in the bridging sequence.  The dashed grey lines indicate the
benchmark estimate ($\pm\,$SE) from Neal (\citeyear{nea99}), and the $c$ values
of the horizontal axis refer to the design of the partial data
posterior bridging sequence as $r_j =
n_\mathrm{tot}\times\{0,1/9,2/9,\ldots,1\}^{c}$. These sequences are
also illustrated graphically for clarity in the right-hand panel.}\label{root}
\end{figure*}

\begin{figure*}[b]

\includegraphics{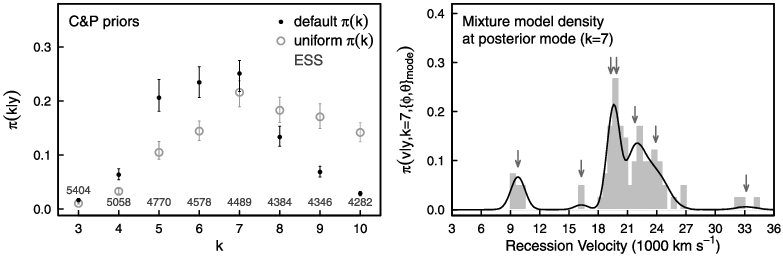}

\caption{Posterior probabilities for the number of Normal mixture
components in the  galaxy data set, $\pi(k|y)$, under our
astronomically motivated priors (left-hand panel).  The solid, dark grey symbols here
denote the true posterior, while the open, light grey symbols indicate for
reference the raw, \textup{``}per component\textup{''} marginal
likelihood-based result, that is, before application of our truncated Poisson
$\pi(k)$. In each case the relevant uncertainties
[recovered from estimates of the asymptotic covariance matrix for
each $\log \hat{Z}^{(k)}$]
are illustrated as 95\% credible interval error bars.  The effective
sample size (ESS) provided by the pseudo-mixture of 10 partial data
posteriors  sampled for 4000 draws each is noted in grey for each $k$. The
inferred probability density (in velocity space) at the
maximum a posteriori parameterization of our Normal mixture model ($k=7$) is
then illustrated for reference against a scaled histogram of the galaxy data set in the right-hand panel.}
\label{galaxypost}
\end{figure*}

\subsubsection{Infinite mixture model}

The same considerations can also
help shape our hyperparameter choices for the priors on our infinite
mixture model. In particular, we take
$\{m=17,s=4,h_1=2,h_2=8,\nu_1=1,\allowbreak  \nu_2=1\}$ for the hyperparameters shaping the
Normal-Gamma reference density, $G_0$, with the aim of matching as
closely as possible to the priors of our finite-dimensional model.  With the
scale parameter of our prior on the component precisions taking an
inverse-Gamma hyperprior form in the infinite case and a
Gamma form in the finite case, it was not possible to  exactly match these distributions: our choice of $\{\nu_1=1,\nu_2=1\}$
is intended to at least give comparable 5\% and 95\% quantiles.  Finally,
we adopt a fixed value for the concentration parameter of $M=1$;
this choice coincidentally gives a similar effective prior for the number
of unique components among the 82 observed galaxies to that of the
$\pi(k)$ adopted for our finite mixture model (see \citealt{esc95}, for instance).

\begin{figure*}

\includegraphics{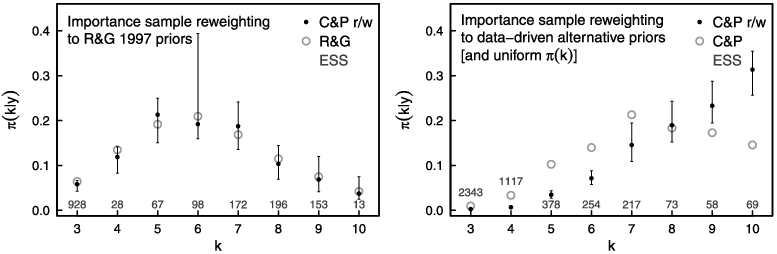}

\caption{Importance sample reweighting of our
draws from the pseudo-mixture of partial data posteriors used to estimate
$\pi(k|y)$ under the Richardson and Green (\citeyear{ric97}) priors (left-hand panel).  The solid, dark grey symbols here
denote the reweighted posterior estimate, while the open, light grey symbols indicate for
reference the Richardson and Green (\citeyear{ric97}) benchmark.  The results of the equivalent
procedure to approximate the effect of using alternative
\textup{``}data-driven\textup{''} priors are shown in the right-hand panel; here the solid,
dark grey symbols again represent the reweighted estimate, with the
open, light grey symbols illustrating the reference point provided by our
default priors.  In both instances we treat $\pi(k)$ as uniform to
emphasize the difference to the \textup{``}per component\textup{''} marginal
likelihoods made by this modest change of prior.  In each panel the relevant uncertainties
[recovered via bootstrap resampling from our pseudo-mixture plus estimates of the asymptotic covariance matrix for
each $\log \hat{Z}^{(k)}$]
are illustrated as 95\% credible interval error bars.  The effective
sample size (ESS) provided by the pseudo-mixture of 10 partial data
posteriors sampled for 4000 draws each is noted in grey for each $k$.}
\label{priorsensitivity}\vspace*{-3pt}
\end{figure*}

\begin{figure*}[b]

\includegraphics{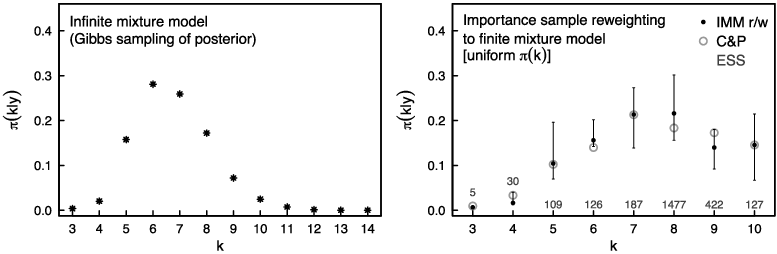}

\caption{Posterior for the number of unique label
assignments among the galaxy data set recovered from Gibbs
sampling of our (Dirichlet process-based) infinite mixture model
(left-hand panel).  The results of importance sample reweighting of
these draws, combined as a pseudo-mixture with those simulated under
our bridging sequence of partial data posteriors, are shown via the solid, dark grey symbols in the
right-hand panel.  The target of this reweighting procedure is the \textup{``}per
component\textup{''} [i.e., uniform $\pi(k)$] posterior for the number of
mixture model components in our benchmark finite mixture model
(shown as the open, light grey symbols). The relevant uncertainties
[recovered via bootstrap resampling from our pseudo-mixture plus estimates of the asymptotic covariance matrix for
each $\log \hat{Z}^{(k)}$]
are illustrated as 95\% credible interval error bars.  The effective
sample size (ESS) provided by the pseudo-mixture of 10 partial data
posteriors sampled for 4000 draws each is noted in grey for each $k$.}
\label{imm}
\end{figure*}

\subsection{Numerical Results}\label{priorsense}

\subsubsection{Chib example}

As an initial verification of our code, we first run the Gibbs sampling
procedure outlined above (Section~\ref{fmm}) to explore the partial
data posteriors of a three-component (unequal variance)
mixture model using the priors from
\citet{chi95}, with the biased sampling algorithm then applied for marginal likelihood estimation.  \citet{nea99} has made public the
results of a $10^8$ draw AME calculation providing a precise benchmark for the marginal
likelihood under these priors of $-226.791\ (\pm \,0.089)$ (SE), though
it should be noted that the galaxy data set used for this purpose is
that \textit{with} Chib's transcription error in the 78th observation (which we
insert explicitly into the public \texttt{R} version
for the present application only).  Given just 200 saved draws from
Gibbs sampling (at a thinning rate of 0.9) of the partial data
posterior at each of 10
steps spaced as $r_j = \lfloor
n_\mathrm{tot}\times\{0,1/9,2/9,\ldots,1,\}^{c=2} \rfloor$ (with $r_2$ reset
to 3 to facilitate sampling), we can confirm the recovery of this benchmark as
$-226.79\ (\pm \,0.15)$ (SE).  Estimation of the (single run) standard error (SE) was for
this purpose conducted via 1000 repeat simulations.  Further repeats of this
procedure with both more posterior focused ($c=0.5$, 1) and more prior
focused ($c=4$) partial data schedules confirm the optimality of
the $c=2$ choice anticipated from Fisher information principles
(Section~\ref{fisher}).  The results of this experiment are illustrated in Figure~\ref{root}.

\subsubsection{Finite mixture model}

To estimate ``per component'' marginal likelihoods for each $k$ ($3
\le k \le 10$) in our finite mixture model, we run the same procedure
of partial data posterior exploration followed by biased sampling
with $4000$ draws from each of ten steps on the $c=2$ bridging
sequence.  The results of this computation are illustrated in
Figure~\ref{galaxypost}; the uncertainties indicated are gauged\vadjust{\goodbreak} from the asymptotic covariance
matrix of the biased sampling estimator (as per \citealt{gil88}).  We
recover a posterior mode of $k=7$ components, the recession velocity
density belonging to
which at the corresponding mode in $\{\phi,\theta\}$ is also
illustrated in Figure~\ref{galaxypost} for reference.  To the eye, it
appears that $k=7$ may be a slight overestimate since the third and
fourth components (in order of increasing recession velocity) are
more or less on top of each other, suggesting that one is being used
to account for a slight non-Normality in the shape of this peak.

To demonstrate the potential for efficient prior-sensitivity analysis
via importance sample reweighting of the pseudo-mixture density of
partial data posteriors
normalized by biased sampling (Section~\ref{psa}), we begin by recovering the \citet{ric97}
result from the above simulation output.  The results of this
reweighting procedure are shown in Figure~\ref{priorsensitivity}.
Since the \citet{ric97} priors are significantly different to those
chosen here from astronomical considerations (as discussed in Section~\ref{hyperp}), the effective sample sizes
provided by our pseudo-mixture of $4000 \times 10$ draws range from
just 13 to 928, yet the resulting approximation to the former
benchmark is actually rather good.  Moreover, the corresponding
95\% credible intervals [recovered via bootstrap resampling from our pseudo-mixture plus estimates of the asymptotic covariance matrix for
each $\log \hat{Z}^{(k)}$] indeed enclose all eight $\pi(k|y)$ reference
points.

As a second demonstration we also show in Figure~\ref{priorsensitivity}
the results of reweighting for alternative
``data-driven'' choices for the hyperparameters of our prior on the
component means: $\{\kappa=\bar{y}\ (20.8),\xi=1/\operatorname{var}\{y\}\allowbreak   (0.048)\}$. To emphasize the large
difference this small change in $\pi(\theta)$ makes to the ``per component'' log
$Z^{(k)}$ values, the comparison presented is between our default and
``data-driven'' priors with $\pi(k)$ removed (i.e., treated as
uniform).  This investigation clearly confirms the
remarkable prior sensitivity of $\pi(k|y)$ in the galaxy data set
example.  Interestingly, the preference under our ``data-driven'' priors
is for an even greater number of mixture components ($k > 7$), despite
the $k=7$ solution already seeming (visually) to be an overfitting of the available data.\vadjust{\goodbreak}

\subsubsection{Infinite mixture model}\label{immr}

In Figure~\ref{imm} we present the results of Gibbs sampling the
posterior of our infinite mixture model.  In particular, we
show in the left-hand panel of this figure the posterior for the number
of unique label assignments among the galaxy data set, which, as we
have noted earlier, is typically used as a proxy for the number of
mixture components present (although under the Dirichet process prior
this is formally always
infinite).  In the right-hand panel we demonstrate again the power of
importance sample reweighting for prior-sensitivity analysis, though
for this particular case the stochastic process prior used requires
that we apply the appropriate Radon--Nikodym derivative version given
by Equation (\ref{rnreweighting}).

The Radon--Nikodym derivative, $\frac{dP_{\{z,\theta,\psi\},\mathrm{alt}}}{dP_{\{z,\theta,\psi\}}}(\{z,\theta,\allowbreak  \psi\})$, of the measure on
$\{z,\theta,  \psi\}$ assigned by a $k$-component finite mixture
 model with respect to that assigned by the Dirichlet process
prior of our
infinite mixture model may be computed as follows.  First, we observe
that the Radon--Nikodym derivative between two Dirichlet process priors
on the (equivalent) space of
$\{\{\theta_1,\ldots,\theta_{n_\mathrm{tot}}\},\psi\}$ (with the
$\theta_i$ possibly nonunique) has been previously derived by \citet{dos12}, thereby providing a direct
formula for computing
$\frac{dP_{\{z,\theta,\psi\},\mathrm{int}}}{dP_{\{z,\theta,\psi\}}}(\{z,\theta,\psi\})$, where
$P_{\{z,\theta,\psi\},\mathrm{int}}$ represents a Dirichlet process prior with
hyperpriors on the $\psi$ of its reference density chosen to be identical to those on the
$\{\mu_j,\tau_j\}$ and $\beta$ of our finite mixture model.  That
is, we choose $P_{\{z,\theta,\psi\},\mathrm{int}}$ such that its
projection to $P_{\{z,\theta,\psi\},\mathrm{int}}$ for $z$ with $k$
unique elements is equivalent (a.e.) to that of
$P_{\{z,\theta\},\mathrm{alt}}$ with our hyperparameter on $\beta$
integrated out, allowing $P_{\{z,\theta,\psi\},\mathrm{alt}}$ to
be defined identical to $f(z)P_{\{z,\theta,\psi\},\mathrm{int}}$.
The necessary $f(z)$ to ensure that
$\frac{dP_{\{z,\theta,\psi\},\mathrm{alt}}}{dP_{\{z,\theta,\psi\},\mathrm{int}}}\frac{dP_{\{z,\theta,\psi\},\mathrm{int}}}{dP_{\{z,\theta,\psi\}}}
=
\frac{dP_{\{z,\theta,\psi\},\mathrm{alt}}}{dP_{\{z,\theta,\psi\}}}$
is then simply the ratio of the labeling probabilities under our finite
mixture model and the intermediate version of our infinite mixture
model [with $f(z)\neq 0$ \textit{only} where the number of unique elements in $z$
equals~$k$].

A formula for the desired $f(z)$ can be derived by combining
standard properties of the Dirichlet-Multinomial distribution (our
finite-dimensional model
prior on $z$) with results from the work of \citet{ant74} on the
marginals of the Dirichlet process.  In each case the probability
of a given labeling sequence depends not on its ordering, but rather
on its vector of per-label counts.  Using Antoniak's system
of writing $C(m_1,m_2,\ldots,m_{n_\mathrm{tot}})$ as the set of labelings with
$m_1$ unique elements, $m_2$ pairs, etc., we have wherever
$\sum_{i=1}^{n_\mathrm{tot}}m_i \le k$,
\begin{eqnarray*}
f(z \in C) &=& \Biggl(\frac{
n_\mathrm{tot}!}{\prod_{i=1}^{n_\mathrm{tot}}(i!)^{m_i}} \frac{\Gamma(k\alpha)}{\Gamma(n_\mathrm{tot}+k\alpha)}
\\
&&\phantom{\qquad\quad}{}\cdot\prod
_{i=1}^{n_\mathrm{tot}} \biggl(\frac{\Gamma(i+\alpha)}{\Gamma(\alpha)}
\biggr)^{m_i}\Biggr)
\\
&&{}\Big/ \biggl(\frac{(k-\sum_{i=1}^{n_\mathrm{tot}}m_i)!(\prod m_i)!}{k!}
\\
&&\hspace*{0.5cm}{}\cdot\frac{n_\mathrm{tot}!}{\prod_{i=1}^{n_\mathrm{tot}}
i^{m_i} (m_i!)} \frac{M^{\sum_{i=1}^{n_\mathrm{tot}}m_i}}{M^{[n_\mathrm{tot}]}}\biggr),
\end{eqnarray*}
where
$x^{[j]}$ denotes the rising factorial function
as per Proposition~3 of \citet{ant74}.  For our case of $\alpha=1$
and $M=1$ this reduces to
\begin{eqnarray*}
f(z \in C) &=& \biggl( \frac{n_\mathrm{tot}!(k-1)!}{(n_\mathrm{tot}+k-1)!} \biggr)
\\
&&{}\Big/ \biggl(\frac{(k-\sum_{i=1}^{n_\mathrm{tot}}m_i)!(\prod m_i)!}{k! \prod_{i=1}^{n_\mathrm{tot}}
i^{m_i} (m_i!)}
\biggr).
\end{eqnarray*}

\section{Conclusions}\label{conclusions}

In this paper we have presented an extensive review of the recursive
pathway to marginal likelihood estimation as characterized by
biased
sampling, reverse logistic regression and the density of states; in
particular, we have highlighted the diversity of bridging sequences
amenable to recursive normalization and the utility of the resulting
pseudo-mixtures for prior-sensitivity analysis (in the Bayesian
context).  Our key theoretical contributions have included the
introduction of a novel heuristic (``thermodynamic integration via
importance sampling'') for guiding design of the bridging sequence
and an elucidation of various connections between these recursive
estimators and the nested sampling technique.  Our two numerical case
studies illustrate in depth the practical implementation of these
ideas using both ``ordinary'' and stochastic process priors.




\begin{thebibliography}{91}

\bibitem[\protect\citeauthoryear{Aitkin}{2001}]{ait01}
\begin{barticle}[auto:STB|2014/02/12|12:18:25]
\bauthor{\bsnm{Aitkin},~\bfnm{M.}\binits{M.}}
(\byear{2001}).
\btitle{Likelihood and Bayesian analysis of mixtures}.
\bjournal{Statist. Model.}
\bvolume{1}
\bpages{287--304}.
\end{barticle}
\bptok{imsref}%
\endbibitem

\bibitem[\protect\citeauthoryear{Antoniak}{1974}]{ant74}
\begin{barticle}[mr]
\bauthor{\bsnm{Antoniak},~\bfnm{Charles~E.}\binits{C.~E.}}
(\byear{1974}).
\btitle{Mixtures of {D}irichlet processes with applications to {B}ayesian nonparametric problems}.
\bjournal{Ann. Statist.}
\bvolume{2}
\bpages{1152--1174}.
\bid{issn={0090-5364}, mr={0365969}}
\end{barticle}
\bptok{imsref}%
\endbibitem

\bibitem[\protect\citeauthoryear{Arima and Tardella}{2012}]{ari12}
\begin{barticle}[mr]
\bauthor{\bsnm{Arima},~\bfnm{Serena}\binits{S.}} \AND
\bauthor{\bsnm{Tardella},~\bfnm{Luca}\binits{L.}}
(\byear{2012}).
\btitle{Improved harmonic mean estimator for phylogenetic model evidence}.
\bjournal{J. Comput. Biol.}
\bvolume{19}
\bpages{418--438}.
\bid{doi={10.1089/cmb.2010.0139}, issn={1066-5277}, mr={2913981}}
\end{barticle}
\bptok{imsref}%
\endbibitem

\bibitem[\protect\citeauthoryear{Baele et~al.}{2012}]{bae12}
\begin{barticle}[pbm]
\bauthor{\bsnm{Baele},~\bfnm{Guy}\binits{G.}},
\bauthor{\bsnm{Lemey},~\bfnm{Philippe}\binits{P.}},
\bauthor{\bsnm{Bedford},~\bfnm{Trevor}\binits{T.}},
\bauthor{\bsnm{Rambaut},~\bfnm{Andrew}\binits{A.}},
\bauthor{\bsnm{Suchard},~\bfnm{Marc~A.}\binits{M.~A.}} \AND
\bauthor{\bsnm{Alekseyenko},~\bfnm{Alexander~V.}\binits{A.~V.}}
(\byear{2012}).
\btitle{Improving the accuracy of demographic and molecular clock model comparison while accommodating phylogenetic uncertainty}.
\bjournal{Mol. Biol. Evol.}
\bvolume{29}
\bpages{2157--2167}.
\bid{doi={10.1093/molbev/mss084}, issn={1537-1719}, pii={mss084}, pmcid={3424409}, pmid={22403239}}
\end{barticle}
\bptok{imsref}%
\endbibitem

\bibitem[\protect\citeauthoryear{Bailer-Jones}{2012}]{bai12}
\begin{barticle}[auto:STB|2014/02/12|12:18:25]
\bauthor{\bsnm{Bailer-Jones},~\bfnm{C.~A.~L.}\binits{C.~A.~L.}}
(\byear{2012}).
\btitle{A Bayesian method for the analysis of deterministic and stochastic time series}.
\bjournal{Astron. Astrophys.}
\bvolume{546}
\bpages{A89}.
\end{barticle}
\bptok{imsref}%
\endbibitem

\bibitem[\protect\citeauthoryear{Billingsley}{1968}]{bil68}
\begin{bbook}[mr]
\bauthor{\bsnm{Billingsley},~\bfnm{Patrick}\binits{P.}}
(\byear{1968}).
\btitle{Convergence of Probability Measures}.
\bpublisher{Wiley},
\blocation{New York}.
\bid{mr={0233396}}
\end{bbook}
\bptok{imsref}%
\endbibitem

\bibitem[\protect\citeauthoryear{Brewer and Stello}{2009}]{bre09}
\begin{barticle}[auto:STB|2014/02/12|12:18:25]
\bauthor{\bsnm{Brewer},~\bfnm{B.~J.}\binits{B.~J.}} \AND
\bauthor{\bsnm{Stello},~\bfnm{D.}\binits{D.}}
(\byear{2009}).
\btitle{Gaussian process modelling of asteroseismic data}.
\bjournal{Mon. Not. R. Astron. Soc.}
\bvolume{395}
\bpages{2226--2233}.
\end{barticle}
\bptok{imsref}%
\endbibitem

\bibitem[\protect\citeauthoryear{Caimo and Friel}{2013}]{cai12}
\begin{barticle}[auto:STB|2014/02/12|12:18:25]
\bauthor{\bsnm{Caimo},~\bfnm{A.}\binits{A.}} \AND
\bauthor{\bsnm{Friel},~\bfnm{N.}\binits{N.}}
(\byear{2013}).
\btitle{Bayesian model selection for exponential random graph models}.
\bjournal{Social Networks}
\bvolume{35}
\bpages{11--24}.
\end{barticle}
\bptok{imsref}%
\endbibitem

\bibitem[\protect\citeauthoryear{Calderhead and Girolami}{2009}]{cal09}
\begin{barticle}[mr]
\bauthor{\bsnm{Calderhead},~\bfnm{Ben}\binits{B.}} \AND
\bauthor{\bsnm{Girolami},~\bfnm{Mark}\binits{M.}}
(\byear{2009}).
\btitle{Estimating {B}ayes factors via thermodynamic integration and population {MCMC}}.
\bjournal{Comput. Statist. Data Anal.}
\bvolume{53}
\bpages{4028--4045}.
\bid{doi={10.1016/j.csda.2009.07.025}, issn={0167-9473}, mr={2744303}}
\end{barticle}
\bptok{imsref}%
\endbibitem

\bibitem[\protect\citeauthoryear{Cameron and Pettitt}{2013}]{cam13}
\begin{bmisc}[auto:STB|2014/02/12|12:18:25]
\bauthor{\bsnm{Cameron},~\bfnm{E.}\binits{E.}} \AND
\bauthor{\bsnm{Pettitt},~\bfnm{A.~N.}\binits{A.~N.}}
(\byear{2013}).
\bhowpublished{On the evidence for cosmic variation of the fine structure constant (II):
A~semi-parametric Bayesian model selection analysis of the quasar dataset. Preprint. Available at \arxivurl{arXiv:1309.2737}}.
\end{bmisc}
\bptok{imsref}%
\endbibitem

\bibitem[\protect\citeauthoryear{Capp{\'e} et~al.}{2004}]{cap04}
\begin{barticle}[mr]
\bauthor{\bsnm{Capp{\'e}},~\bfnm{O.}\binits{O.}},
\bauthor{\bsnm{Guillin},~\bfnm{A.}\binits{A.}},
\bauthor{\bsnm{Marin},~\bfnm{J.~M.}\binits{J.~M.}} \AND
\bauthor{\bsnm{Robert},~\bfnm{C.~P.}\binits{C.~P.}}
(\byear{2004}).
\btitle{Population {M}onte {C}arlo}.
\bjournal{J. Comput. Graph. Statist.}
\bvolume{13}
\bpages{907--929}.
\bid{doi={10.1198/106186004X12803}, issn={1061-8600}, mr={2109057}}
\end{barticle}
\bptok{imsref}%
\endbibitem

\bibitem[\protect\citeauthoryear{Chen and Shao}{1997}]{che97}
\begin{barticle}[mr]
\bauthor{\bsnm{Chen},~\bfnm{Ming-Hui}\binits{M.-H.}} \AND
\bauthor{\bsnm{Shao},~\bfnm{Qi-Man}\binits{Q.-M.}}
(\byear{1997}).
\btitle{On {M}onte {C}arlo methods for estimating ratios of normalizing constants}.
\bjournal{Ann. Statist.}
\bvolume{25}
\bpages{1563--1594}.
\bid{doi={10.1214/aos/1031594732}, issn={0090-5364}, mr={1463565}}
\end{barticle}
\bptok{imsref}%
\endbibitem

\bibitem[\protect\citeauthoryear{Chen, Shao and Ibrahim}{2000}]{che00}
\begin{bbook}[mr]
\bauthor{\bsnm{Chen},~\bfnm{Ming-Hui}\binits{M.-H.}},
\bauthor{\bsnm{Shao},~\bfnm{Qi-Man}\binits{Q.-M.}} \AND
\bauthor{\bsnm{Ibrahim},~\bfnm{Joseph~G.}\binits{J.~G.}}
(\byear{2000}).
\btitle{Monte {C}arlo Methods in {B}ayesian Computation}.
\bpublisher{Springer},
\blocation{New York}.
\bid{doi={10.1007/978-1-4612-1276-8}, mr={1742311}}
\end{bbook}
\bptok{imsref}%
\endbibitem

\bibitem[\protect\citeauthoryear{Chib}{1995}]{chi95}
\begin{barticle}[mr]
\bauthor{\bsnm{Chib},~\bfnm{Siddhartha}\binits{S.}}
(\byear{1995}).
\btitle{Marginal likelihood from the {G}ibbs output}.
\bjournal{J.~Amer. Statist. Assoc.}
\bvolume{90}
\bpages{1313--1321}.
\bid{issn={0162-1459}, mr={1379473}}
\end{barticle}
\bptok{imsref}%
\endbibitem

\bibitem[\protect\citeauthoryear{Chopin}{2002}]{cho02}
\begin{barticle}[mr]
\bauthor{\bsnm{Chopin},~\bfnm{Nicolas}\binits{N.}}
(\byear{2002}).
\btitle{A sequential particle filter method for static models}.
\bjournal{Biometrika}
\bvolume{89}
\bpages{539--551}.
\bid{doi={10.1093/biomet/89.3.539}, issn={0006-3444}, mr={1929161}}
\end{barticle}
\bptok{imsref}%
\endbibitem

\bibitem[\protect\citeauthoryear{Chopin and Robert}{2010}]{cho10}
\begin{barticle}[mr]
\bauthor{\bsnm{Chopin},~\bfnm{Nicolas}\binits{N.}} \AND
\bauthor{\bsnm{Robert},~\bfnm{Christian~P.}\binits{C.~P.}}
(\byear{2010}).
\btitle{Properties of nested sampling}.
\bjournal{Biometrika}
\bvolume{97}
\bpages{741--755}.
\bid{doi={10.1093/biomet/asq021}, issn={0006-3444}, mr={2672495}}
\end{barticle}
\bptok{imsref}%
\endbibitem

\bibitem[\protect\citeauthoryear{Cornuet et~al.}{2012}]{cor12}
\begin{barticle}[mr]
\bauthor{\bsnm{Cornuet},~\bfnm{Jean-Marie}\binits{J.-M.}},
\bauthor{\bsnm{Marin},~\bfnm{Jean-Michel}\binits{J.-M.}},
\bauthor{\bsnm{Mira},~\bfnm{Antonietta}\binits{A.}} \AND
\bauthor{\bsnm{Robert},~\bfnm{Christian~P.}\binits{C.~P.}}
(\byear{2012}).
\btitle{Adaptive multiple importance sampling}.
\bjournal{Scand. J. Stat.}
\bvolume{39}
\bpages{798--812}.
\bid{doi={10.1111/j.1467-9469.2011.00756.x}, issn={0303-6898}, mr={3000850}}
\end{barticle}
\bptok{imsref}%
\endbibitem

\bibitem[\protect\citeauthoryear{Davis et al.}{2007}]{dav07}
\begin{barticle}[auto:STB|2014/02/12|12:18:25]
\bauthor{\bsnm{Davis},~\bfnm{T. M.}\binits{T. M.}} \betal{et~al.}
(\byear{2007}).
\btitle{Scrutinizing exotic cosmological models using ESSENCE supernova data combined with other cosmological probes}.
\bjournal{Astrophys. J.}
\bvolume{666}
\bpages{716--725}.
\end{barticle}
\bptok{imsref}%
\endbibitem

\bibitem[\protect\citeauthoryear{Del~Moral, Doucet and Jasra}{2006}]{del06}
\begin{barticle}[mr]
\bauthor{\bsnm{Del Moral},~\bfnm{Pierre}\binits{P.}},
\bauthor{\bsnm{Doucet},~\bfnm{Arnaud}\binits{A.}} \AND
\bauthor{\bsnm{Jasra},~\bfnm{Ajay}\binits{A.}}
(\byear{2006}).
\btitle{Sequential {M}onte {C}arlo samplers}.
\bjournal{J. R. Stat. Soc. Ser. B Stat. Methodol.}
\bvolume{68}
\bpages{411--436}.
\bid{doi={10.1111/j.1467-9868.2006.00553.x}, issn={1369-7412}, mr={2278333}}
\end{barticle}
\bptok{imsref}%
\endbibitem

\bibitem[\protect\citeauthoryear{Diebolt and Robert}{1994}]{die94}
\begin{barticle}[mr]
\bauthor{\bsnm{Diebolt},~\bfnm{Jean}\binits{J.}} \AND
\bauthor{\bsnm{Robert},~\bfnm{Christian~P.}\binits{C.~P.}}
(\byear{1994}).
\btitle{Estimation of finite mixture distributions through {B}ayesian sampling}.
\bjournal{J. Roy. Statist. Soc. Ser. B}
\bvolume{56}
\bpages{363--375}.
\bid{issn={0035-9246}, mr={1281940}}
\end{barticle}
\bptok{imsref}%
\endbibitem

\bibitem[\protect\citeauthoryear{Doss}{2012}]{dos12}
\begin{barticle}[mr]
\bauthor{\bsnm{Doss},~\bfnm{Hani}\binits{H.}}
(\byear{2012}).
\btitle{Hyperparameter and model selection for nonparametric {B}ayes problems via {R}adon--{N}ikodym derivatives}.
\bjournal{Statist. Sinica}
\bvolume{22}
\bpages{1--26}.
\bid{doi={10.5705/ss.2009.259}, issn={1017-0405}, mr={2933165}}
\end{barticle}
\bptok{imsref}%
\endbibitem

\bibitem[\protect\citeauthoryear{Dudley and Philipp}{1983}]{dud83}
\begin{barticle}[mr]
\bauthor{\bsnm{Dudley},~\bfnm{R.~M.}\binits{R.~M.}} \AND
\bauthor{\bsnm{Philipp},~\bfnm{Walter}\binits{W.}}
(\byear{1983}).
\btitle{Invariance principles for sums of {B}anach space valued random elements and empirical processes}.
\bjournal{Z. Wahrsch. Verw. Gebiete}
\bvolume{62}
\bpages{509--552}.
\bid{doi={10.1007/BF00534202}, issn={0044-3719}, mr={0690575}}
\end{barticle}
\bptok{imsref}%
\endbibitem

\bibitem[\protect\citeauthoryear{Escobar and West}{1995}]{esc95}
\begin{barticle}[mr]
\bauthor{\bsnm{Escobar},~\bfnm{Michael~D.}\binits{M.~D.}} \AND
\bauthor{\bsnm{West},~\bfnm{Mike}\binits{M.}}
(\byear{1995}).
\btitle{Bayesian density estimation and inference using mixtures}.
\bjournal{J. Amer. Statist. Assoc.}
\bvolume{90}
\bpages{577--588}.
\bid{issn={0162-1459}, mr={1340510}}
\end{barticle}
\bptok{imsref}%
\endbibitem

\bibitem[\protect\citeauthoryear{Evans et~al.}{2003}]{jia03}
\begin{barticle}[auto:STB|2014/02/12|12:18:25]
\bauthor{\bsnm{Evans},~\bfnm{M.}\binits{M.}},
\bauthor{\bsnm{Robert},~\bfnm{C.~P.}\binits{C.~P.}},
\bauthor{\bsnm{Davison},~\bfnm{A.~C.}\binits{A.~C.}},
\bauthor{\bsnm{Jiang},~\bfnm{W.}\binits{W.}},
\bauthor{\bsnm{Tanner},~\bfnm{M.~A.}\binits{M.~A.}},
\bauthor{\bsnm{Doss},~\bfnm{H.}\binits{H.}},
\bauthor{\bsnm{Qin},~\bfnm{J.}\binits{J.}},
\bauthor{\bsnm{Fokianos},~\bfnm{K.}\binits{K.}},
\bauthor{\bsnm{MacEachern},~\bfnm{S.~N.}\binits{S.~N.}},
\bauthor{\bsnm{Peruggia},~\bfnm{M.}\binits{M.}},
\bauthor{\bsnm{Guha},~\bfnm{S.}\binits{S.}},
\bauthor{\bsnm{Chib},~\bfnm{S.}\binits{S.}},
\bauthor{\bsnm{Ritov},~\bfnm{Y.}\binits{Y.}},
\bauthor{\bsnm{Robins},~\bfnm{J.~M.}\binits{J.~M.}} \AND
\bauthor{\bsnm{Vardi},~\bfnm{Y.}\binits{Y.}}
(\byear{2003}).
\btitle{Discussion on the paper by Kong, McCullagh, Meng, Nicolas and Tan}.
\bjournal{J. Roy. Statist. Soc. B}
\bvolume{65}
\bpages{604--618}.
\end{barticle}
\bptok{imsref}%
\endbibitem

\bibitem[\protect\citeauthoryear{Fan et~al.}{2012}]{fan12}
\begin{barticle}[auto:STB|2014/02/12|12:18:25]
\bauthor{\bsnm{Fan},~\bfnm{Y.}\binits{Y.}},
\bauthor{\bsnm{Rui},~\bfnm{W.}\binits{W.}},
\bauthor{\bsnm{Chen},~\bfnm{M.-H.}\binits{M.-H.}},
\bauthor{\bsnm{Kuo},~\bfnm{L.}\binits{L.}} \AND
\bauthor{\bsnm{Lewis},~\bfnm{P.~O.}\binits{P.~O.}}
(\byear{2012}).
\btitle{Choosing among partition models in Bayesian phylogenetics}.
\bjournal{Mol. Biol. Evol.}
\bvolume{28}
\bpages{523--532}.
\end{barticle}
\bptok{imsref}%
\endbibitem

\bibitem[\protect\citeauthoryear{Feroz and Hobson}{2008}]{fer08}
\begin{barticle}[auto:STB|2014/02/12|12:18:25]
\bauthor{\bsnm{Feroz},~\bfnm{F.}\binits{F.}} \AND
\bauthor{\bsnm{Hobson},~\bfnm{M.~P.}\binits{M.~P.}}
(\byear{2008}).
\btitle{Multimodal nested sampling: An efficient and robust alternative to Markov Chain Monte Carlo methods for astronomical data analyses}.
\bjournal{Mon. Not. R. Astron. Soc.}
\bvolume{384}
\bpages{449--463}.
\end{barticle}
\bptok{imsref}%
\endbibitem

\bibitem[\protect\citeauthoryear{Feroz et~al.}{2013}]{fer13}
\begin{bmisc}[auto:STB|2014/02/12|12:18:25]
\bauthor{\bsnm{Feroz},~\bfnm{F.}\binits{F.}},
\bauthor{\bsnm{Hobson},~\bfnm{M.~P.}\binits{M.~P.}},
\bauthor{\bsnm{Cameron},~\bfnm{E.}\binits{E.}} \AND
\bauthor{\bsnm{Pettitt},~\bfnm{A.~N.}\binits{A.~N.}}
(\byear{2013}).
\bhowpublished{Importance nested sampling and the multinest algorithm. Preprint. Available at
\arxivurl{arXiv:1306.2144}}.
\end{bmisc}
\bptok{imsref}%
\endbibitem

\bibitem[\protect\citeauthoryear{Ferrenberg and Swendsen}{1989}]{fer89}
\begin{barticle}[pbm]
\bauthor{\bsnm{Ferrenberg},~\bfnm{A.~M.}\binits{A.~M.}} \AND
\bauthor{\bsnm{Swendsen},~\bfnm{R.~H.}\binits{R.~H.}}
(\byear{1989}).
\btitle{Optimized Monte Carlo data analysis}.
\bjournal{Phys. Rev. Lett.}
\bvolume{63}
\bpages{1195--1198}.
\bid{issn={1079-7114}, pmid={10040500}}
\end{barticle}
\bptok{imsref}%
\endbibitem

\bibitem[\protect\citeauthoryear{Friel and Pettitt}{2008}]{fri08}
\begin{barticle}[mr]
\bauthor{\bsnm{Friel},~\bfnm{N.}\binits{N.}} \AND
\bauthor{\bsnm{Pettitt},~\bfnm{A.~N.}\binits{A.~N.}}
(\byear{2008}).
\btitle{Marginal likelihood estimation via power posteriors}.
\bjournal{J. R. Stat. Soc. Ser. B Stat. Methodol.}
\bvolume{70}
\bpages{589--607}.
\bid{doi={10.1111/j.1467-9868.2007.00650.x}, issn={1369-7412}, mr={2420416}}
\end{barticle}
\bptok{imsref}%
\endbibitem

\bibitem[\protect\citeauthoryear{Friel and Wyse}{2012}]{fri12}
\begin{barticle}[mr]
\bauthor{\bsnm{Friel},~\bfnm{Nial}\binits{N.}} \AND
\bauthor{\bsnm{Wyse},~\bfnm{Jason}\binits{J.}}
(\byear{2012}).
\btitle{Estimating the evidence---A~review}.
\bjournal{Stat. Neerl.}
\bvolume{66}
\bpages{288--308}.
\bid{doi={10.1111/j.1467-9574.2011.00515.x}, issn={0039-0402}, mr={2955421}}
\end{barticle}
\bptok{imsref}%
\endbibitem

\bibitem[\protect\citeauthoryear{Gelfand and Dey}{1994}]{gel94}
\begin{barticle}[mr]
\bauthor{\bsnm{Gelfand},~\bfnm{A.~E.}\binits{A.~E.}} \AND
\bauthor{\bsnm{Dey},~\bfnm{D.~K.}\binits{D.~K.}}
(\byear{1994}).
\btitle{Bayesian model choice: Asymptotics and exact calculations}.
\bjournal{J. Roy. Statist. Soc. Ser. B}
\bvolume{56}
\bpages{501--514}.
\bid{issn={0035-9246}, mr={1278223}}
\end{barticle}
\bptok{imsref}%
\endbibitem

\bibitem[\protect\citeauthoryear{Gelman and Meng}{1998}]{gel98}
\begin{barticle}[mr]
\bauthor{\bsnm{Gelman},~\bfnm{Andrew}\binits{A.}} \AND
\bauthor{\bsnm{Meng},~\bfnm{Xiao-Li}\binits{X.-L.}}
(\byear{1998}).
\btitle{Simulating normalizing constants: From importance sampling to bridge sampling to path sampling}.
\bjournal{Statist. Sci.}
\bvolume{13}
\bpages{163--185}.
\bid{doi={10.1214/ss/1028905934}, issn={0883-4237}, mr={1647507}}
\end{barticle}
\bptok{imsref}%
\endbibitem

\bibitem[\protect\citeauthoryear{Gelman et~al.}{2004}]{gel04}
\begin{bbook}[mr]
\bauthor{\bsnm{Gelman},~\bfnm{Andrew}\binits{A.}},
\bauthor{\bsnm{Carlin},~\bfnm{John~B.}\binits{J.~B.}},
\bauthor{\bsnm{Stern},~\bfnm{Hal~S.}\binits{H.~S.}} \AND
\bauthor{\bsnm{Rubin},~\bfnm{Donald~B.}\binits{D.~B.}}
(\byear{2004}).
\btitle{Bayesian Data Analysis},
\bedition{2nd} ed.
\bpublisher{Chapman \& Hall/CRC},
\blocation{Boca Raton, FL}.
\bid{mr={2027492}}
\end{bbook}
\bptok{imsref}%
\endbibitem

\bibitem[\protect\citeauthoryear{Geyer}{1992}]{gey92b}
\begin{barticle}[auto:STB|2014/02/12|12:18:25]
\bauthor{\bsnm{Geyer},~\bfnm{C.~J.}\binits{C.~J.}}
(\byear{1992}).
\btitle{Practical Markov chain Monte Carlo}.
\bjournal{Statist. Sci.}
\bvolume{7}
\bpages{473--483}.
\end{barticle}
\bptok{imsref}%
\endbibitem

\bibitem[\protect\citeauthoryear{Geyer}{1994}]{gey94}
\begin{bmisc}[auto:STB|2014/02/12|12:18:25]
\bauthor{\bsnm{Geyer},~\bfnm{C.~J.}\binits{C.~J.}}
(\byear{1994}).
\bhowpublished{Estimating normalizing constants and reweighting mixtures in Markov chain Monte Carlo. Technical Report 568, School of Statistics,
Univ. Minnesota, Minneapolis, MN}.
\end{bmisc}
\bptok{imsref}%
\endbibitem

\bibitem[\protect\citeauthoryear{Geyer and Thompson}{1992}]{gey92}
\begin{barticle}[mr]
\bauthor{\bsnm{Geyer},~\bfnm{Charles~J.}\binits{C.~J.}} \AND
\bauthor{\bsnm{Thompson},~\bfnm{Elizabeth~A.}\binits{E.~A.}}
(\byear{1992}).
\btitle{Constrained {M}onte {C}arlo maximum likelihood for dependent data}.
\bjournal{J. Roy. Statist. Soc. Ser. B}
\bvolume{54}
\bpages{657--699}.
\bid{issn={0035-9246}, mr={1185217}}
\bptnote{check related}%
\end{barticle}
\bptok{imsref}%
\endbibitem

\bibitem[\protect\citeauthoryear{Gill, Vardi and Wellner}{1988}]{gil88}
\begin{barticle}[mr]
\bauthor{\bsnm{Gill},~\bfnm{Richard~D.}\binits{R.~D.}},
\bauthor{\bsnm{Vardi},~\bfnm{Yehuda}\binits{Y.}} \AND
\bauthor{\bsnm{Wellner},~\bfnm{Jon~A.}\binits{J.~A.}}
(\byear{1988}).
\btitle{Large sample theory of empirical distributions in biased sampling models}.
\bjournal{Ann. Statist.}
\bvolume{16}
\bpages{1069--1112}.
\bid{doi={10.1214/aos/1176350948}, issn={0090-5364}, mr={0959189}}
\end{barticle}
\bptok{imsref}%
\endbibitem

\bibitem[\protect\citeauthoryear{Gr\"{u}n and Leisch}{2010}]{gru10}
\begin{bmisc}[auto:STB|2014/02/12|12:18:25]
\bauthor{\bsnm{Gr\"{u}n},~\bfnm{B.}\binits{B.}} \AND
\bauthor{\bsnm{Leisch},~\bfnm{F.}\binits{F.}}
(\byear{2010}).
\bhowpublished{BayesMix: An R package for Bayesian mixture modeling. Technical report}.
\end{bmisc}
\bptok{imsref}%
\endbibitem

\bibitem[\protect\citeauthoryear{Gunn}{1972}]{gun72}
\begin{barticle}[auto:STB|2014/02/12|12:18:25]
\bauthor{\bsnm{Gunn},~\bfnm{J.~E.}\binits{J.~E.}} \AND
\bauthor{\bsnm{Gott},~\bfnm{J. R. III}\binits{J. R. III}}
(\byear{1972}).
\btitle{On the infall of matter into clusters of galaxies and some effects on their evolution}.
\bjournal{Astrophys. J.}
\bvolume{176}
\bpages{1--19}.
\end{barticle}
\bptok{imsref}%
\endbibitem

\bibitem[\protect\citeauthoryear{Habeck}{2012}]{hab12}
\begin{barticle}[auto:STB|2014/02/12|12:18:25]
\bauthor{\bsnm{Habeck},~\bfnm{M.}\binits{M.}}
(\byear{2012}).
\btitle{Evaluation of marginal likelihoods via the density of states}.
\bjournal{J. Mach. Learn. Res.}
\bvolume{22}
\bpages{486--494}.
\end{barticle}
\bptok{imsref}%
\endbibitem

\bibitem[\protect\citeauthoryear{Halmos}{1950}]{hal50}
\begin{bbook}[mr]
\bauthor{\bsnm{Halmos},~\bfnm{Paul~R.}\binits{P.~R.}}
(\byear{1950}).
\btitle{Measure {T}heory}.
\bpublisher{Van Nostrand},
\blocation{New York}.
\bid{mr={0033869}}
\end{bbook}
\bptok{imsref}%
\endbibitem

\bibitem[\protect\citeauthoryear{Halmos and Savage}{1949}]{hal49}
\begin{barticle}[mr]
\bauthor{\bsnm{Halmos},~\bfnm{Paul~R.}\binits{P.~R.}} \AND
\bauthor{\bsnm{Savage},~\bfnm{L.~J.}\binits{L.~J.}}
(\byear{1949}).
\btitle{Application of the {R}adon--{N}ikodym theorem to the theory of sufficient statistics}.
\bjournal{Ann. Math. Statist.}
\bvolume{20}
\bpages{225--241}.
\bid{issn={0003-4851}, mr={0030730}}
\end{barticle}
\bptok{imsref}%
\endbibitem

\bibitem[\protect\citeauthoryear{Hesterberg}{1995}]{hes91}
\begin{barticle}[auto:STB|2014/02/12|12:18:25]
\bauthor{\bsnm{Hesterberg},~\bfnm{T.}\binits{T.}}
(\byear{1995}).
\btitle{Weighted average importance sampling and defensive mixture distributions}.
\bjournal{Technometrics}
\bvolume{37}
\bpages{185--194}.
\end{barticle}
\bptok{imsref}%
\endbibitem

\bibitem[\protect\citeauthoryear{Hoeting et~al.}{1999}]{hoe99}
\begin{barticle}[mr]
\bauthor{\bsnm{Hoeting},~\bfnm{Jennifer~A.}\binits{J.~A.}},
\bauthor{\bsnm{Madigan},~\bfnm{David}\binits{D.}},
\bauthor{\bsnm{Raftery},~\bfnm{Adrian~E.}\binits{A.~E.}} \AND
\bauthor{\bsnm{Volinsky},~\bfnm{Chris~T.}\binits{C.~T.}}
(\byear{1999}).
\btitle{Bayesian model averaging: A tutorial}.
\bjournal{Statist. Sci.}
\bvolume{14}
\bpages{382--417}.
\bid{doi={10.1214/ss/1009212519}, issn={0883-4237}, mr={1765176}}
\end{barticle}
\bptok{imsref}%
\endbibitem

\bibitem[\protect\citeauthoryear{H{\"o}rmander}{1983}]{hor83}
\begin{bbook}[mr]
\bauthor{\bsnm{H{\"o}rmander},~\bfnm{Lars}\binits{L.}}
(\byear{1983}).
\btitle{The Analysis of Linear Partial Differential Operators. {I}: Distribution Theory and Fourier Analysis}.
\bseries{Grundlehren der Mathematischen Wissenschaften [Fundamental Principles of Mathematical Sciences]}
\bvolume{256}.
\bpublisher{Springer},
\blocation{Berlin}.
\bid{doi={10.1007/978-3-642-96750-4}, mr={0717035}}
\end{bbook}
\bptok{imsref}%
\endbibitem


\bibitem[\protect\citeauthoryear{Jara et~al.}{2011}]{jar11}
\begin{barticle}[auto:STB|2014/02/12|12:18:25]
\bauthor{\bsnm{Jara},~\bfnm{A.}\binits{A.}},
\bauthor{\bsnm{Hanson},~\bfnm{T.~E.}\binits{T.~E.}},
\bauthor{\bsnm{Quintana},~\bfnm{F.~A.}\binits{F.~A.}},
\bauthor{\bsnm{M{\"u}ller},~\bfnm{P.}\binits{P.}} \AND
\bauthor{\bsnm{Rosner},~\bfnm{G.~L.}\binits{G.~L.}}
(\byear{2011}).
\btitle{DPpackage: Bayesian semi- and nonparametric modelling in R}.
\bjournal{J. Statist. Softw.}
\bvolume{40}
\bpages{1--30}.
\end{barticle}
\bptok{imsref}%
\endbibitem

\bibitem[\protect\citeauthoryear{Jasra, Holmes and Stephens}{2005}]{jas05}
\begin{barticle}[mr]
\bauthor{\bsnm{Jasra},~\bfnm{A.}\binits{A.}},
\bauthor{\bsnm{Holmes},~\bfnm{C.~C.}\binits{C.~C.}} \AND
\bauthor{\bsnm{Stephens},~\bfnm{D.~A.}\binits{D.~A.}}
(\byear{2005}).
\btitle{Markov chain {M}onte {C}arlo methods and the label switching problem in {B}ayesian mixture modeling}.
\bjournal{Statist. Sci.}
\bvolume{20}
\bpages{50--67}.
\bid{doi={10.1214/088342305000000016}, issn={0883-4237}, mr={2182987}}
\end{barticle}
\bptok{imsref}%
\endbibitem

\bibitem[\protect\citeauthoryear{Jaynes}{2003}]{jay03}
\begin{bbook}[mr]
\bauthor{\bsnm{Jaynes},~\bfnm{E.~T.}\binits{E.~T.}}
(\byear{2003}).
\btitle{Probability Theory: The Logic of Science}.
\bpublisher{Cambridge Univ. Press},
\blocation{Cambridge}.
\bid{doi={10.1017/CBO9780511790423}, mr={1992316}}
\end{bbook}
\bptok{imsref}%
\endbibitem

\bibitem[\protect\citeauthoryear{Jeffreys}{1961}]{jef61}
\begin{bbook}[mr]
\bauthor{\bsnm{Jeffreys},~\bfnm{Harold}\binits{H.}}
(\byear{1961}).
\btitle{Theory of Probability},
\bedition{3rd} ed.
\bpublisher{Clarendon Press},
\blocation{Oxford}.
\bid{mr={0187257}}
\end{bbook}
\bptok{imsref}%
\endbibitem

\bibitem[\protect\citeauthoryear{Jeffreys and Berger}{1991}]{jef91}
\begin{bmisc}[auto:STB|2014/02/12|12:18:25]
\bauthor{\bsnm{Jeffreys},~\bfnm{W.~H.}\binits{W.~H.}} \AND
\bauthor{\bsnm{Berger},~\bfnm{J.~O.}\binits{J.~O.}}
(\byear{1991}).
\bhowpublished{Sharpening Ockham's razor on a Bayesian strop. Technical Report 91-44C, Dept. Statistics,
Purdue Univ., West Lafayette, IN}.
\end{bmisc}
\bptok{imsref}%
\endbibitem

\bibitem[\protect\citeauthoryear{Kass and Raftery}{1995}]{kas95}
\begin{barticle}[auto:STB|2014/02/12|12:18:25]
\bauthor{\bsnm{Kass},~\bfnm{R.~E.}\binits{R.~E.}} \AND
\bauthor{\bsnm{Raftery},~\bfnm{A.~E.}\binits{A.~E.}}
(\byear{1995}).
\btitle{Bayes factors}.
\bjournal{J. Amer. Statist. Assoc.}
\bvolume{90}
\bpages{773--795}.
\end{barticle}
\bptok{imsref}%
\endbibitem\vadjust{\goodbreak}

\bibitem[\protect\citeauthoryear{Kilbinger et~al.}{2010}]{kil10}
\begin{barticle}[auto:STB|2014/02/12|12:18:25]
\bauthor{\bsnm{Kilbinger},~\bfnm{M.}\binits{M.}},
\bauthor{\bsnm{Wraith},~\bfnm{D.}\binits{D.}},
\bauthor{\bsnm{Robert},~\bfnm{C.~P.}\binits{C.~P.}},
\bauthor{\bsnm{Benabed},~\bfnm{K.}\binits{K.}},
\bauthor{\bsnm{Capp{\'e}},~\bfnm{O.}\binits{O.}},
\bauthor{\bsnm{Cardoso},~\bfnm{J.-F.}\binits{J.-F.}},
\bauthor{\bsnm{Fort},~\bfnm{G.}\binits{G.}},
\bauthor{\bsnm{Prunet},~\bfnm{S.}\binits{S.}} \AND
\bauthor{\bsnm{Bouchet},~\bfnm{F.~R.}\binits{F.~R.}}
(\byear{2010}).
\btitle{Bayesian model comparison in cosmology with population Monte Carlo}.
\bjournal{Mon. Not. R. Astron. Soc.}
\bvolume{405}
\bpages{2381--2390}.
\end{barticle}
\bptok{imsref}%
\endbibitem

\bibitem[\protect\citeauthoryear{Kirshner et~al.}{1981}]{kir81}
\begin{barticle}[auto:STB|2014/02/12|12:18:25]
\bauthor{\bsnm{Kirshner},~\bfnm{R.~P.}\binits{R.~P.}},
\bauthor{\bsnm{Oemler},~\bfnm{A.~Jr.}\binits{A.~Jr.}},
\bauthor{\bsnm{Schechter},~\bfnm{P.~L.}\binits{P.~L.}} \AND
\bauthor{\bsnm{Shectman},~\bfnm{S.~A.}\binits{S.~A.}}
(\byear{1981}).
\btitle{A million cubic megaparsec void in Bo\"otes?}
\bjournal{Astrophys. J.}
\bvolume{248}
\bpages{57--60}.
\end{barticle}
\bptok{imsref}%
\endbibitem

\bibitem[\protect\citeauthoryear{Kong, Liu and Wong}{1994}]{kon94}
\begin{barticle}[auto:STB|2014/02/12|12:18:25]
\bauthor{\bsnm{Kong},~\bfnm{A.}\binits{A.}},
\bauthor{\bsnm{Liu},~\bfnm{J.~S.}\binits{J.~S.}} \AND
\bauthor{\bsnm{Wong},~\bfnm{W.~H.}\binits{W.~H.}}
(\byear{1994}).
\btitle{Sequential imputations and Bayesian missing data problems}.
\bjournal{J. Amer. Statist. Assoc.}
\bvolume{89}
\bpages{278--288}.
\end{barticle}
\bptok{imsref}%
\endbibitem

\bibitem[\protect\citeauthoryear{Kong et~al.}{2003}]{kon03}
\begin{barticle}[mr]
\bauthor{\bsnm{Kong},~\bfnm{A.}\binits{A.}},
\bauthor{\bsnm{McCullagh},~\bfnm{P.}\binits{P.}},
\bauthor{\bsnm{Meng},~\bfnm{X.-L.}\binits{X.-L.}},
\bauthor{\bsnm{Nicolae},~\bfnm{D.}\binits{D.}} \AND
\bauthor{\bsnm{Tan},~\bfnm{Z.}\binits{Z.}}
(\byear{2003}).
\btitle{A theory of statistical models for {M}onte {C}arlo integration}.
\bjournal{J. R. Stat. Soc. Ser. B Stat. Methodol.}
\bvolume{65}
\bpages{585--618}.
\bid{doi={10.1111/1467-9868.00404}, issn={1369-7412}, mr={1998624}}
\bptnote{check related}%
\end{barticle}
\bptok{imsref}%
\endbibitem

\bibitem[\protect\citeauthoryear{Kumar et~al.}{1992}]{kum92}
\begin{barticle}[auto:STB|2014/02/12|12:18:25]
\bauthor{\bsnm{Kumar},~\bfnm{S.}\binits{S.}},
\bauthor{\bsnm{Rosenberg},~\bfnm{J.~M.}\binits{J.~M.}},
\bauthor{\bsnm{Bouzida},~\bfnm{D.}\binits{D.}},
\bauthor{\bsnm{Swendsen},~\bfnm{R.~H.}\binits{R.~H.}} \AND
\bauthor{\bsnm{Kollman},~\bfnm{P.~A.}\binits{P.~A.}}
(\byear{1992}).
\btitle{The weighted~histogram analysis method for free-energy calculations on~bio\-molecules. I. The method}.
\bjournal{J. Comput. Chem.}
\bvolume{13}
\bpages{1011--1021}.
\end{barticle}
\bptok{imsref}%
\endbibitem

\bibitem[\protect\citeauthoryear{Lartillot and Phillipe}{2006}]{lar06}
\begin{barticle}[auto:STB|2014/02/12|12:18:25]
\bauthor{\bsnm{Lartillot},~\bfnm{N.}\binits{N.}} \AND
\bauthor{\bsnm{Phillipe},~\bfnm{H.}\binits{H.}}
(\byear{2006}).
\btitle{Computing Bayes factors using thermodynamic integration}.
\bjournal{Syst. Biol.}
\bvolume{55}
\bpages{195--207}.
\end{barticle}
\bptok{imsref}%
\endbibitem

\bibitem[\protect\citeauthoryear{Lee et~al.}{2008}]{lee08}
\begin{bmisc}[auto:STB|2014/02/12|12:18:25]
\bauthor{\bsnm{Lee},~\bfnm{K.}\binits{K.}},
\bauthor{\bsnm{Marin},~\bfnm{J.-M.}\binits{J.-M.}},
\bauthor{\bsnm{Mengersen},~\bfnm{K.}\binits{K.}} \AND
\bauthor{\bsnm{Robert},~\bfnm{C.~P.}\binits{C.~P.}}
(\byear{2008}).
\bhowpublished{Bayesian inference on mixtures of distributions. Preprint. Available at \arxivurl{arXiv:0804.2413}}.
\end{bmisc}
\bptok{imsref}%
\endbibitem

\bibitem[\protect\citeauthoryear{Lefebvre, Steele and Vandal}{2010}]{lef10}
\begin{barticle}[mr]
\bauthor{\bsnm{Lefebvre},~\bfnm{Genevi{\`e}ve}\binits{G.}},
\bauthor{\bsnm{Steele},~\bfnm{Russell}\binits{R.}} \AND
\bauthor{\bsnm{Vandal},~\bfnm{Alain~C.}\binits{A.~C.}}
(\byear{2010}).
\btitle{A~path sampling identity for computing the {K}ullback--{L}eibler and {J} divergences}.
\bjournal{Comput. Statist. Data Anal.}
\bvolume{54}
\bpages{1719--1731}.
\bid{doi={10.1016/j.csda.2010.01.018}, issn={0167-9473}, mr={2608968}}
\end{barticle}
\bptok{imsref}%
\endbibitem

\bibitem[\protect\citeauthoryear{Li, Ni and Lin}{2011}]{li11}
\begin{barticle}[mr]
\bauthor{\bsnm{Li},~\bfnm{Yong}\binits{Y.}},
\bauthor{\bsnm{Ni},~\bfnm{Zhong-Xin}\binits{Z.-X.}} \AND
\bauthor{\bsnm{Lin},~\bfnm{Jin-Guan}\binits{J.-G.}}
(\byear{2011}).
\btitle{A stochastic simulation approach to model selection for stochastic volatility models}.
\bjournal{Comm. Statist. Simulation Comput.}
\bvolume{40}
\bpages{1043--1056}.
\bid{doi={10.1080/03610918.2011.563002}, issn={0361-0918}, mr={2792481}}
\end{barticle}
\bptok{imsref}%
\endbibitem

\bibitem[\protect\citeauthoryear{Liu}{2001}]{liu01}
\begin{bbook}[mr]
\bauthor{\bsnm{Liu},~\bfnm{Jun~S.}\binits{J.~S.}}
(\byear{2001}).
\btitle{Monte {C}arlo Strategies in Scientific Computing}.
\bpublisher{Springer},
\blocation{New York}.
\bid{mr={1842342}}
\end{bbook}
\bptok{imsref}%
\endbibitem

\bibitem[\protect\citeauthoryear{Marin, Pudlo and Sedki}{2012}]{mar12}
\begin{bmisc}[author]
\bauthor{\bsnm{Marin},~\bfnm{J.-M.}\binits{J.-M.}},
\bauthor{\bsnm{Pudlo},~\bfnm{P.}\binits{P.}} \AND
\bauthor{\bsnm{Sedki},~\bfnm{M.}\binits{M.}}
(\byear{2012}).
\bhowpublished{Consistency of the adaptive
multiple importance sampling. Preprint. Available at
\arxivurl{arXiv:1301.2548}}.
\end{bmisc}
\bptok{imsref}%
\endbibitem

\bibitem[\protect\citeauthoryear{Marin and Robert}{2010}]{mar10}
\begin{barticle}[mr]
\bauthor{\bsnm{Marin},~\bfnm{Jean-Michel}\binits{J.-M.}} \AND
\bauthor{\bsnm{Robert},~\bfnm{Christian~P.}\binits{C.~P.}}
(\byear{2010}).
\btitle{On resolving the {S}avage--{D}ickey paradox}.
\bjournal{Electron. J. Stat.}
\bvolume{4}
\bpages{643--654}.
\bid{doi={10.1214/10-EJS564}, issn={1935-7524}, mr={2660536}}
\end{barticle}
\bptok{imsref}%
\endbibitem



\bibitem[\protect\citeauthoryear{Meng and Wong}{1996}]{men96}
\begin{barticle}[mr]
\bauthor{\bsnm{Meng},~\bfnm{Xiao-Li}\binits{X.-L.}} \AND
\bauthor{\bsnm{Wong},~\bfnm{Wing~Hung}\binits{W.~H.}}
(\byear{1996}).
\btitle{Simulating ratios of normalizing constants via a simple identity: A theoretical exploration}.
\bjournal{Statist. Sinica}
\bvolume{6}
\bpages{831--860}.
\bid{issn={1017-0405}, mr={1422406}}
\end{barticle}
\bptok{imsref}%
\endbibitem

\bibitem[\protect\citeauthoryear{Miller and Harrison}{2013}]{mil13}
\begin{bmisc}[author]
\bauthor{\bsnm{Miller},~\bfnm{Jeffrey~W.}\binits{J.~W.}} \AND
\bauthor{\bsnm{Harrison},~\bfnm{Matthew~T.}\binits{M.~T.}}
(\byear{2013}).
\bhowpublished{A simple example of Dirichlet
process mixture inconsistency for the number of components. Preprint.
Available at \arxivurl{arXiv:1301.2708v1}}.
\end{bmisc}
\bptok{imsref}%
\endbibitem

\bibitem[\protect\citeauthoryear{Mukherjee, Parkinson and Liddle}{2006}]{muk06}
\begin{barticle}[auto:STB|2014/02/12|12:18:25]
\bauthor{\bsnm{Mukherjee},~\bfnm{P.}\binits{P.}},
\bauthor{\bsnm{Parkinson},~\bfnm{D.}\binits{D.}} \AND
\bauthor{\bsnm{Liddle},~\bfnm{A.~R.}\binits{A.~R.}}
(\byear{2006}).
\btitle{A~nested sampling algorithm for cosmological model selection}.
\bjournal{Astrophys. J.}
\bvolume{638}
\bpages{L51--L54}.
\end{barticle}
\bptok{imsref}%
\endbibitem

\bibitem[\protect\citeauthoryear{Neal}{1999}]{nea99}
\begin{bmisc}[auto:STB|2014/02/12|12:18:25]
\bauthor{\bsnm{Neal},~\bfnm{R.}\binits{R.}}
(\byear{1999}).
\bhowpublished{Erroneous results in
``Marginal likelihood from the Gibbs output.''
Available at \url{http://www.cs.toronto.edu/\textasciitilde radford/ftp/chib-letter.pdf}}.
\end{bmisc}
\bptok{imsref}%
\endbibitem

\bibitem[\protect\citeauthoryear{Neal}{2000}]{nea00}
\begin{barticle}[mr]
\bauthor{\bsnm{Neal},~\bfnm{Radford~M.}\binits{R.~M.}}
(\byear{2000}).
\btitle{Markov chain sampling methods for {D}irichlet process mixture models}.
\bjournal{J. Comput. Graph. Statist.}
\bvolume{9}
\bpages{249--265}.
\bid{doi={10.2307/1390653}, issn={1061-8600}, mr={1823804}}
\end{barticle}
\bptok{imsref}%
\endbibitem

\bibitem[\protect\citeauthoryear{Neal}{2001}]{nea01}
\begin{barticle}[mr]
\bauthor{\bsnm{Neal},~\bfnm{Radford~M.}\binits{R.~M.}}
(\byear{2001}).
\btitle{Annealed importance sampling}.
\bjournal{Stat. Comput.}
\bvolume{11}
\bpages{125--139}.
\bid{doi={10.1023/A:1008923215028}, issn={0960-3174}, mr={1837132}}
\end{barticle}
\bptok{imsref}%
\endbibitem

\bibitem[\protect\citeauthoryear{Newton and Raftery}{1994}]{new94}
\begin{barticle}[mr]
\bauthor{\bsnm{Newton},~\bfnm{Michael~A.}\binits{M.~A.}} \AND
\bauthor{\bsnm{Raftery},~\bfnm{Adrian~E.}\binits{A.~E.}}
(\byear{1994}).
\btitle{Approximate {B}ayesian inference with the weighted likelihood bootstrap}.
\bjournal{J.~Roy. Statist. Soc. Ser. B}
\bvolume{56}
\bpages{3--48}.
\bid{issn={0035-9246}, mr={1257793}}
\bptnote{check related}%
\end{barticle}
\bptok{imsref}%
\endbibitem\vadjust{\goodbreak}

\bibitem[\protect\citeauthoryear{Ortega and Rheinboldt}{1967}]{ort67}
\begin{barticle}[mr]
\bauthor{\bsnm{Ortega},~\bfnm{James~M.}\binits{J.~M.}} \AND
\bauthor{\bsnm{Rheinboldt},~\bfnm{Werner~C.}\binits{W.~C.}}
(\byear{1967}).
\btitle{Monotone iterations for nonlinear equations with application to {G}auss--{S}eidel methods}.
\bjournal{SIAM J. Numer. Anal.}
\bvolume{4}
\bpages{171--190}.
\bid{issn={0036-1429}, mr={0215487}}\vadjust{\vspace*{-1pt}\eject}
\end{barticle}
\bptok{imsref}%
\endbibitem

\bibitem[\protect\citeauthoryear{Pfanzagl}{1979}]{pfa79}
\begin{barticle}[mr]
\bauthor{\bsnm{Pfanzagl},~\bfnm{J.}\binits{J.}}
(\byear{1979}).
\btitle{Conditional distributions as derivatives}.
\bjournal{Ann. Probab.}
\bvolume{7}
\bpages{1046--1050}.
\bid{issn={0091-1798}, mr={0548898}}
\end{barticle}
\bptok{imsref}%
\endbibitem

\bibitem[\protect\citeauthoryear{Phillips and Smith}{1996}]{phi96}
\begin{bincollection}[mr]
\bauthor{\bsnm{Phillips},~\bfnm{David~B.}\binits{D.~B.}} \AND
\bauthor{\bsnm{Smith},~\bfnm{Adrian~F.~M.}\binits{A.~F.~M.}}
(\byear{1996}).
\btitle{Bayesian model comparison via jump diffusions}.
In \bbooktitle{Markov Chain {M}onte {C}arlo in Practice}
\bpages{215--239}.
\bpublisher{Chapman \& Hall},
\blocation{London}.
\bid{mr={1397970}}
\end{bincollection}
\bptok{imsref}%
\endbibitem

\bibitem[\protect\citeauthoryear{Postman, Huchra and Geller}{1986}]{pos86}
\begin{barticle}[auto:STB|2014/02/12|12:18:25]
\bauthor{\bsnm{Postman},~\bfnm{M.}\binits{M.}},
\bauthor{\bsnm{Huchra},~\bfnm{J.~P.}\binits{J.~P.}} \AND
\bauthor{\bsnm{Geller},~\bfnm{M.~J.}\binits{M.~J.}}
(\byear{1986}).
\btitle{Probes of large-scale structure in the Corona Borealis region}.
\bjournal{Astrophys. J.}
\bvolume{92}
\bpages{1238--1247}.
\end{barticle}
\bptok{imsref}%
\endbibitem

\bibitem[\protect\citeauthoryear{Raftery et~al.}{2007}]{raf07}
\begin{bincollection}[mr]
\bauthor{\bsnm{Raftery},~\bfnm{Adrian~E.}\binits{A.~E.}},
\bauthor{\bsnm{Newton},~\bfnm{Michael~A.}\binits{M.~A.}},
\bauthor{\bsnm{Satagopan},~\bfnm{Jaya~M.}\binits{J.~M.}} \AND
\bauthor{\bsnm{Krivitsky},~\bfnm{Pavel~N.}\binits{P.~N.}}
(\byear{2007}).
\btitle{Estimating the integrated likelihood via posterior simulation using the harmonic mean identity}.
In \bbooktitle{Bayesian Statistics 8}
\bpages{371--416}.
\bpublisher{Oxford Univ. Press},
\blocation{Oxford}.
\bid{mr={2433201}}
\end{bincollection}
\bptok{imsref}%
\endbibitem

\bibitem[\protect\citeauthoryear{Richardson and Green}{1997}]{ric97}
\begin{barticle}[mr]
\bauthor{\bsnm{Richardson},~\bfnm{Sylvia}\binits{S.}} \AND
\bauthor{\bsnm{Green},~\bfnm{Peter~J.}\binits{P.~J.}}
(\byear{1997}).
\btitle{On {B}ayesian analysis of mixtures with an unknown number of components}.
\bjournal{J. Roy. Statist. Soc. Ser. B}
\bvolume{59}
\bpages{731--792}.
\bid{doi={10.1111/1467-9868.00095}, issn={0035-9246}, mr={1483213}}
\bptnote{check related, check year}%
\end{barticle}
\bptok{imsref}%
\endbibitem

\bibitem[\protect\citeauthoryear{Robert and Wraith}{2009}]{rob09}
\begin{bincollection}[auto:STB|2014/02/12|12:18:25]
\bauthor{\bsnm{Robert},~\bfnm{C.~P.}\binits{C.~P.}} \AND
\bauthor{\bsnm{Wraith},~\bfnm{D.}\binits{D.}}
(\byear{2009}).
\btitle{Computational methods for Bayesian model choice}.
In \bbooktitle{Bayesian Inference and Maximum Entropy Methods in Science and Engineering: The 29th International Workshop on Bayesian Inference and Maximum Entropy Methods in Science and Engineering}.
\bseries{AIP Conference Proceedings}
\bvolume{1193}
\bpages{251--262}.
\bpublisher{American Institute of Physics},
\baddress{New York}.
\end{bincollection}
\bptok{imsref}%
\endbibitem

\bibitem[\protect\citeauthoryear{Roeder}{1990}]{roe90}
\begin{barticle}[auto:STB|2014/02/12|12:18:25]
\bauthor{\bsnm{Roeder},~\bfnm{K.}\binits{K.}}
(\byear{1990}).
\btitle{Density estimation with confidence sets exemplified by superclusters and voids in the galaxies}.
\bjournal{J. Amer. Statist. Assoc.}
\bvolume{85}
\bpages{617--624}.
\end{barticle}
\bptok{imsref}%
\endbibitem

\bibitem[\protect\citeauthoryear{Roeder and Wasserman}{1997}]{roe97}
\begin{barticle}[mr]
\bauthor{\bsnm{Roeder},~\bfnm{Kathryn}\binits{K.}} \AND
\bauthor{\bsnm{Wasserman},~\bfnm{Larry}\binits{L.}}
(\byear{1997}).
\btitle{Practical {B}ayesian density estimation using mixtures of normals}.
\bjournal{J. Amer. Statist. Assoc.}
\bvolume{92}
\bpages{894--902}.
\bid{doi={10.2307/2965553}, issn={0162-1459}, mr={1482121}}
\end{barticle}
\bptok{imsref}%
\endbibitem

\bibitem[\protect\citeauthoryear{Schechter}{1976}]{sch76}
\begin{barticle}[auto:STB|2014/02/12|12:18:25]
\bauthor{\bsnm{Schechter},~\bfnm{P.}\binits{P.}}
(\byear{1976}).
\btitle{An analytic expression for the luminosity function of galaxies}.
\bjournal{Astrophys. J.}
\bvolume{203}
\bpages{297--306}.
\end{barticle}
\bptok{imsref}%
\endbibitem

\bibitem[\protect\citeauthoryear{Shirts and Chodera}{2008}]{shi08}
\begin{barticle}[auto:STB|2014/02/12|12:18:25]
\bauthor{\bsnm{Shirts},~\bfnm{M.~R.}\binits{M.~R.}} \AND
\bauthor{\bsnm{Chodera},~\bfnm{J.~D.}\binits{J.~D.}}
(\byear{2008}).
\btitle{Statistically optimal analysis of samples from multiple equilibrium states}.
\bjournal{J. Chem. Phys.}
\bvolume{129}
\bpages{124105}.
\end{barticle}
\bptok{imsref}%
\endbibitem

\bibitem[\protect\citeauthoryear{Skilling}{2006}]{ski06}
\begin{barticle}[mr]
\bauthor{\bsnm{Skilling},~\bfnm{John}\binits{J.}}
(\byear{2006}).
\btitle{Nested sampling for general {B}ayesian computation}.
\bjournal{Bayesian Anal.}
\bvolume{1}
\bpages{833--859 (electronic)}.
\bid{doi={10.1214/06-BA127}, issn={1931-6690}, mr={2282208}}
\end{barticle}
\bptok{imsref}%
\endbibitem

\bibitem[\protect\citeauthoryear{Stephens}{2000}]{ste00}
\begin{barticle}[mr]
\bauthor{\bsnm{Stephens},~\bfnm{Matthew}\binits{M.}}
(\byear{2000}).
\btitle{Bayesian analysis of mixture models with an unknown number of components---An alternative to reversible jump methods}.
\bjournal{Ann. Statist.}
\bvolume{28}
\bpages{40--74}.
\bid{doi={10.1214/aos/1016120364}, issn={0090-5364}, mr={1762903}}
\end{barticle}
\bptok{imsref}%
\endbibitem

\bibitem[\protect\citeauthoryear{Tan et~al.}{2012}]{tan12}
\begin{barticle}[auto:STB|2014/02/12|12:18:25]
\bauthor{\bsnm{Tan},~\bfnm{Z.}\binits{Z.}},
\bauthor{\bsnm{Gallicchio},~\bfnm{E.}\binits{E.}},
\bauthor{\bsnm{Lapelosa},~\bfnm{M.}\binits{M.}} \AND
\bauthor{\bsnm{Levy},~\bfnm{R.~M.}\binits{R.~M.}}
(\byear{2012}).
\btitle{Theory of binless multi-state free energy estimation with applications to protein-ligand binding}.
\bjournal{J. Chem. Phys.}
\bvolume{136}
\bpages{144102}.
\end{barticle}
\bptok{imsref}%
\endbibitem

\bibitem[\protect\citeauthoryear{Tierney}{1994}]{tie94}
\begin{barticle}[mr]
\bauthor{\bsnm{Tierney},~\bfnm{Luke}\binits{L.}}
(\byear{1994}).
\btitle{Markov chains for exploring posterior distributions}.
\bjournal{Ann. Statist.}
\bvolume{22}
\bpages{1701--1762}.
\bid{doi={10.1214/aos/1176325750}, issn={0090-5364}, mr={1329166}}
\bptnote{check related}%
\end{barticle}
\bptok{imsref}%
\endbibitem

\bibitem[\protect\citeauthoryear{Vardi}{1985}]{var85}
\begin{barticle}[mr]
\bauthor{\bsnm{Vardi},~\bfnm{Y.}\binits{Y.}}
(\byear{1985}).
\btitle{Empirical distributions in selection bias models}.
\bjournal{Ann. Statist.}
\bvolume{13}
\bpages{178--205}.
\bid{doi={10.1214/aos/1176346585}, issn={0090-5364}, mr={0773161}}
\bptnote{check related}%
\end{barticle}
\bptok{imsref}%
\endbibitem

\bibitem[\protect\citeauthoryear{Weinberg}{2012}]{wei12}
\begin{barticle}[mr]
\bauthor{\bsnm{Weinberg},~\bfnm{Martin~D.}\binits{M.~D.}}
(\byear{2012}).
\btitle{Computing the {B}ayes factor from a {M}arkov chain {M}onte {C}arlo simulation of the posterior distribution}.
\bjournal{Bayesian Anal.}
\bvolume{7}
\bpages{737--769}.
\bid{doi={10.1214/12-BA725}, issn={1936-0975}, mr={2981634}}
\end{barticle}
\bptok{imsref}%
\endbibitem

\bibitem[\protect\citeauthoryear{Wolpert and Schmidler}{2012}]{wol12}
\begin{barticle}[mr]
\bauthor{\bsnm{Wolpert},~\bfnm{Robert~L.}\binits{R.~L.}} \AND
\bauthor{\bsnm{Schmidler},~\bfnm{Scott~C.}\binits{S.~C.}}
(\byear{2012}).
\btitle{{$\alpha$}-stable limit laws for harmonic mean estimators of marginal likelihoods}.
\bjournal{Statist. Sinica}
\bvolume{22}
\bpages{1233--1251}.
\bid{doi={10.5705/ss.2010.221}, issn={1017-0405}, mr={2987490}}
\end{barticle}
\bptok{imsref}%
\endbibitem

\bibitem[\protect\citeauthoryear{Xie et~al.}{2011}]{xie11}
\begin{barticle}[auto:STB|2014/02/12|12:18:25]
\bauthor{\bsnm{Xie},~\bfnm{W.}\binits{W.}},
\bauthor{\bsnm{Lewis},~\bfnm{P.}\binits{P.}},
\bauthor{\bsnm{Fan},~\bfnm{Y.}\binits{Y.}},
\bauthor{\bsnm{Kuo},~\bfnm{L.}\binits{L.}} \AND
\bauthor{\bsnm{Chen},~\bfnm{M.-H.}\binits{M.-H.}}
(\byear{2011}).
\btitle{Improving marginal likelihood estimation for Bayesian phylogenetic model selection}.
\bjournal{Syst. Biol.}
\bvolume{18}
\bpages{1001--1013}.
\end{barticle}
\bptok{imsref}%
\endbibitem

\end{thebibliography}
\end{document}